\documentstyle[aas2pp4]{article}

\def\ref{\par\noindent\hangindent=0.6in\hangafter=1}
\def\caII{\hbox{Ca~$\scriptstyle\rm II$}}
\def\caIII{\hbox{Ca~$\scriptstyle\rm III$}}
\def\mgII{\hbox{Mg~$\scriptstyle\rm II$}}

\begin{document}

\title{Chromospheric \caII\ H and K Emission Among Subdwarfs}

\author{Graeme H. Smith and Christopher W. Churchill\altaffilmark{1}}
\affil{University of California Observatories/Lick Observatory, \\
University of California, Santa Cruz, California, 95064 USA}

\altaffiltext{1}{Present address: Department of Astronomy and Astrophysics,
   Pennsylvania State University, University Park, Pennsylvania, 16801 USA}

\begin{abstract}

Echelle spectra have been obtained of the \caII\ H and K lines for a sample
of metal-poor subdwarf stars as well as for a number of nearby 
Population~I dwarfs selected
from among those included in the Mount Wilson HK survey. The main conclusion
of this paper is that \caII\ H- and K-line emission 
does occur among subdwarfs. It is particularly notable
among those subdwarfs with colours of $B-V \geq 0.75$; all such stars
observed exhibit chromospheric emission, although emission is observed among
some subdwarfs bluer than this colour.
The \caII\ K emission profile in most
subdwarfs exhibits an asymmetry of $V/R >1$, similar to that seen in the
integrated light of the solar disk. Two quantitative indicators of the
contrast between the peaks in the emission profile and the neighbouring
photospheric line profile are introduced. Measurements of these indicators
show that the level of \caII\ emission among the subdwarfs is similar to that
among low-activity Population~I dwarfs.

\end{abstract}

\begin{center}
Accepted: {\it MNRAS}
\end{center}

\section{Introduction}

Chromospheric activity among main-sequence stars is well known to be a 
decreasing function of age. The fluxes in the chromospheric 
emission features formed in the cores
of the \caII\ H and K lines show a decrease with age among
stars younger than 4.5 Gyr (Skumanich 1972).  Within most main-sequence
stars chromospheric heating is thought to be governed by the 
activity of an interior magnetic dynamo, the strength of which depends on the 
stellar rotation rate. As main-sequence stars age they spin down,
with a consequent decrease in dynamo activity and chromospheric heating
(Hartmann \& Noyes 1987).

The dynamo model has met with considerable success in describing the
evolution of chromospheric activity among main-sequence stars. However, much 
of the data published to date on chromospheric activity among dwarf
stars of spectral types F-G-K pertains to stars younger than $\sim$ 5 Gyr. 
Such is true, for example, of the studies by Skumanich (1972) and 
Simon, Herbig, \& Boesgaard (1985). By comparison, there is much less 
spectroscopy available pertaining to chromospheric activity among older dwarfs, 
particularly halo subdwarfs. Consequently, little is known about the evolution
of chromospheric activity among the oldest main-sequence stars in the Galaxy. 
To study the low levels of activity expected among such stars,
we have carried out a spectroscopic study of the \caII\ H and K lines for a
sample of low-metallicity subdwarfs. Metal-poor subdwarfs which are
members of the Galactic halo are amongst the
oldest stars in the Galaxy. They are useful for
studying chromospheric activity at great ages. 

The observations of \caII\ H and K line emission among subdwarfs reported in
this paper complement the HST GHRS observations of \mgII\ lines among
metal-poor solar-like stars by Peterson \& Schrijver (1997). 

\section{Observations}

High-resolution spectra of both the \caII\ H and K lines were obtained for a
sample of subdwarfs using the Hamilton echelle spectrograph and the Lick
Observatory 3-m Shane telescope at Mount Hamilton. A TI 800$\times$800 CCD
was used as the detector. Data were obtained during
a number of observing runs: one night in May 1991, one night in September 1991, 
3 nights in September 1994, 2 in November 1994,
and 2 nights in each of April and June 1995.
Observations of many of the stars took the form of several consecutive 
exposures of 2700 sec duration each. 
Total integration times per star typically ranged from 5400 sec to 10,800 sec.
Calibration exposures obtained during the observing runs
included exposures of a ThAr lamp to be used for wavelength calibration of the
data, and exposures of a quartz lamp to be used as flat-field frames. 
Most of the data were reduced using standard IRAF packages, with the
exception that the scattered-light background was subtracted using the
algorithms described by Churchill \& Allen (1995). The spectra that were
obtained in 1991 for HD 25329, HD 103095, HD 157948, and HD 188510, 
were reduced with the VISTA package.

In order to quantify the strength of the \caII\ K emission line relative to 
the photospheric profile, two parameters have been measured from 
spectra obtained in either the 1994 or 1995 observing runs. The first of these
compares the intensity in the two K-line emission peaks to the
photospheric profile on both sides of this emission:
\begin{displaymath}
K = \frac{I_{\rm P1} + I_{\rm P2}}{I_{\rm B1} + I_{\rm B2}}
\end{displaymath}
where $I_{\rm P1}$ and $I_{\rm P2}$ denote the recorded counts in the blueward
and redward peaks respectively of the K-line emission profile, while 
$I_{\rm B1}$ and $I_{\rm B2}$ denote the counts in the local minima 
where the blueward and redward wings of the 
emission profile intersect the photospheric profile. The definitions of
$I_{\rm B1}$, $I_{\rm B2}$, $I_{\rm P1}$, and $I_{\rm P2}$ are illustrated
in Figure~1. 
In some cases where emission is seen more distinctly in the blueward component 
of the emission profile, and only weakly or not at all in the red component, the
parameter $K_{\rm v} = I_{\rm P1}/I_{\rm B1}$ was measured. 
Stars which show either no emission,
or inflections in the core without local emission maxima, are assigned the
designation $K_{\rm v} \leq 1$. Such stars may still have chromospheres as
indicated by the model \caII\ K line spectra of Oranje (1983); small amounts of
network and/or plage emission can be present on the surface of a solar-like
dwarf without producing distinct emission peaks in the line core.
In the case where
$I_{\rm B1} = I_{\rm B2}$, then $K = 0.5 (K_{\rm v} + I_{\rm P2}/I_{\rm B1})$;
if the blueward emission peak is stronger than the redward peak then
$I_{\rm P2} < I_{\rm P1}$ and $K < K_{\rm v}$.

Two samples of stars were observed in this program: (i) a sample of metal-poor
subdwarfs chosen from the lists of Carney \& Latham (1987) and 
Laird, Carney,
\& Latham (1988), and (ii) a number of Population~I dwarfs chosen from Table 1 
of Noyes et~al.~(1984) that were observed as part of the Mount Wilson
HK program (Vaughan \& Preston 1980; Duncan et~al.~1991). The Mount Wilson 
survey provides a large database of \caII\ H and K emission line strengths
for nearby dwarfs, and so constitutes a list of Population~I 
stars suitable for comparison with the subdwarfs. The stars from the 
Mount Wilson survey that were observed with the Shane telescope
are listed in Table~1, together with $V$ magnitudes, $B-V$ colours, spectral
types (mostly from the Bright Star Catalog), and values of the \caII\
chromospheric emission index $S$ from the Mount Wilson HK program.
The majority of the $S$ values were
obtained from the tabulation of Noyes et~al. (1984), except in the case of
HD 184499, for which data are given by Duncan et~al. (1991). 
Also noted in the table (columns 6--9) are the properties of the \caII\ emission
as determined from the Mount Hamilton spectra. Column 6 indicates whether
emission was observed in the \caII\ lines, while column 7 gives an indication 
of the asymmetry of the emission profile in terms of the parameter 
$V/R$ (cf. Wilson 1976), 
which specifies whether the peak in the violet wing of the emission 
profile is greater than ($V/R > 1$), comparable to ($V/R \approx 1$), or less 
than ($V/R < 1$) the peak in the redward wing. The assignment of $V/R$ 
to one of these three categories for each star was done by inspection of 
plots of the spectra.
In addition, the values of the two emission profile parameters,
$K$ and $K_{\rm v}$, measured from the Mount Hamilton spectra,
are listed in columns 8 and 9.
  
The subdwarfs observed in this program are listed in Table~2, together with
photometric and metallicity data obtained mostly from Carney \& Latham (1987), 
and Laird, Carney, \& Latham (1988).
The photometric data for HD 134169 was obtained from Carney (1979).
Column 1 gives the HD or BD designation of the subdwarfs observed. 
The alternative designations from the Lowell Proper Motion Survey 
(Giclas, Burnham, \& Thomas 1971) are given in column 2.
Most of the stars listed in Table~2 have metallicities of [Fe/H] $\leq -0.8$.
There are however some stars of higher metallicity such as HD 221613, HD 224087,
and particularly HD 153344, which may well be old disk dwarfs rather than
halo subdwarfs. The radial velocities of these stars from Carney \& Latham
(1987) are 0.5, $-24$, and $-90$ km/s respectively, which in at least the first
two cases is consistent with old disk membership. The case for HD 153344 is
less clear. Table~2 indicates whether any \caII\ emission lines were apparent 
in the Mount Hamilton spectra. In some cases it is difficult to decide whether
a star exhibits very weak emission or no emission; such cases are designated
with a semi-colon in column 6. In
most cases the H and K emission lines exhibit the same asymmetry; in such cases 
the $V/R$ asymmetry is indicated in column 7. In a few cases where the 
asymmetry of these two emission lines differ, a remark is given in the 
footnotes, which also contain other miscellaneous comments on the character of
the emission. The values of the parameters $K$ and $K_{\rm v}$ are also
listed in Table~2. 

\section{Discussion of the Spectra}

Representative spectra of both the Mount Wilson stars and the subdwarfs 
observed are shown in Figures 2--9. The wavelength scale in these figures
is the rest frame of the ThAr lamp used to obtain the wavelength calibration.
Of the stars observed from the Mount Wilson
survey, two of them, HD 78366 and HD 101501, are classified by Noyes
et~al.~(1984) as very active dwarfs. The spectra of these two stars 
are shown in Figures 2a and 2b. Although both stars exhibit H and K line
emission asymmetries of $V/R > 1$, this asymmetry is not very pronounced; both
the $R$ peaks are very strong and only slightly weaker than the $V$ 
peaks.\footnote{This association of strong emission with nearly-symmetric 
profiles may have a counterpart in the activity of the
Sun. During the solar activity cycle the K and H emission profiles in the
integrated light of the solar disk are most nearly symmetric in active
regions and at the time of solar maximum (White \& Livingston 1981).} 
The other stars observed from the Mount
Wilson survey are classified as relatively low-activity stars. Spectra for
two such stars, HD 115617 and HD 141004, are shown in Figures 3a and 3b. The
differences in \caII\ H and K emission line strengths between the stars in
Figures 2 and 3 are readily apparent. The star HD 190406 has weak emission with
a $V/R$ asymmetry only slightly greater than unity, as shown in Figure 4a.
Most of the stars listed in Table~1 exhibit a \caII\ emission line asymmetry
of $V/R >1$, the same as for the integrated solar disk
(White \& Livingston 1981) and many other dwarf stars  
(cf. Linsky et~al. 1979; Basri \& Linsky 1979). The only star observed 
from the Mount Wilson survey that exhibits a clear asymmetry of $V/R < 1$
is HD 176051. The spectrum of this star is shown in Figure 4b. 

Among some of the bluer stars ($B-V < 0.6$) that were observed from the 
Mount Wilson survey the spectra do not exhibit a 
clear double-peaked K-line emission profile, but instead show
slight inflections or irregularities in slope near the core of the
dominant absorption profile. 
An example of such a star is HD~159332, the spectrum of which is shown in
Figure~5a. In the spectra of such stars there are no local
maxima due to emission within $\pm 1$ \AA\ of the K line core. Another example
of this phenomenon is HD 184499. In the spectrum of this star, shown in Figure
5b, the H line exhibits weak emission, whereas the K line core exhibits
an inflection in slope, becoming steeper near the core. The lack of discernible
emission features in the H and K lines of Population~I dwarfs with
spectral types as early as mid-F to early-F 
has been attributed by Oranje \& Zwaan (1985) to the ionisation of
\caII\ to \caIII\ within their atmospheres.

We can compare our results for the Mount Wilson stars with synthetic \caII\
K line spectra computed by Oranje (1983). Using a mean plage emission profile
derived for the Sun, Oranje (1983) computed model \caII\ K-line-core
profiles for dwarf stars of similar colour to the Sun and a range of fluxes
in the K line core. For dwarfs with K line emission fluxes comparable to those
exhibited by the Sun during the solar cycle, Oranje's model profiles would 
show emission peaks in the K line core with an asymmetry of $V/R > 1$. This is
consistent with the behaviour seen among many of the lower-activity dwarfs
in our Mount Wilson sample. For solar-colour dwarfs with fluxes in the K line 
core lower than occurs at any time during the solar cycle, 
distinct emission peaks are not seen in Oranje's models.
Instead they exhibit inflections or small changes in slope near the line core. 
Such a circumstance is exhibited by HD 184499 (spectral type G0V).
For a star such as this Oranje presumes that the 
surface magnetic flux arises only from a magnetic network and not from plages. 
Oranje's models with the highest K-line-core fluxes have pronounced emission
profiles with $V/R < 1$. By contrast, the two stars from our Mount Wilson sample
with the greatest core fluxes, HD 78366 and HD 101501, 
show emission profiles that are nearly symmetrical. Oranje's models
indicate that the $V/R$ asymmetry of the K line emission profile will change 
from $> 1$ to $< 1$ with increasing amounts of plage emission. It is therefore
likely that the plage coverage in the models could be made to reproduce the
near-symmetry of the profiles of HD 78366 and HD 101501. Such models would have
slightly lower activity levels than Oranje's highest-activity models.

All of the metal-poor dwarfs and subdwarfs having colours of $B-V \geq 0.70$ in 
the sample exhibit \caII\ H and/or K emission.  
The spectra of four of these stars are shown in Figures 6 a--d.
Both the H and K lines of HD 78050, for example (Figure 6c), reveal a 
clear chromospheric emission core in which the violet emission peak 
is slightly stronger than the redward peak. An asymmetry of 
$V/R > 1$ is typical of most subdwarfs that show \caII\ emission.
Among those showing emission is HD 103095 (= Groombridge 1830), 
which is one of the few subdwarfs that were included in the 
Mount Wilson HK survey; it was found from that survey to show an activity 
cycle in H and K emission of period 7.3 yr (Baliunas et~al. 1995).

Unlike stars such as HD 78050, which has a colour of $B-V = 0.80$ and
shows distinct \caII\ H and K emission reversals, a number of the stars in 
the program have colours 
similar to, or bluer than, the Sun, and one (HD 20507) is as blue as
$B-V = 0.45$. The chromospheric K-line emission for such subdwarfs is often 
extremely weak, and for many of the subdwarfs bluer than $B-V = 0.65$
emission has not been conclusively detected.
Since the identification or non-detection of emission for some stars 
might be considered
subjective, particularly for instances of weak emission, plots of the spectra of
three of these subdwarfs are given in Figures 7 a--c. 

Among stars from Table~2 bluer than $B-V = 0.70$, HD~153344 with a colour of 
$B-V = 0.67$ exhibits a clear emission feature in the \caII\ K line. However the
metallicity of this star is [Fe/H] $= -0.3$, and as such it is possibly a 
Galactic disk dwarf and not a Population~II
object.  The bluest subdwarf found to exhibit \caII\ emission is HD~134169
with a colour of $B-V = 0.55$. This subdwarf is comparable in colour to the 
bluest dwarfs from Table~1 that exhibit a K-line emission peak (as opposed to an
inflection near the K line core). The spectrum of this star is shown in
Figure~8. Although there are no clear emission peaks in the core of the H
line, there is a weak emission peak to the violet side of the K-line core.

One star from Table~2 that shows unusual emission profiles is HD 224087, the
spectrum of which is shown in Figure~9. Emission is clearly seen in both the
H and K lines, but the emission profiles, although asymmetric, are 
single-peaked rather than
double-peaked. Carney \& Latham (1987) find this star to be a ``near-certain''
binary on the basis of radial velocity variability.
It is possible that the peculiar activity of this star is related to
it's binarity. 

\section{Emission Profile Parameters}

The largest set of measurements of \caII\ emission line strengths in the 
literature is that resulting from the Mount Wilson program 
(Vaughan \& Preston 1980; Duncan et~al. 1991). The
$S$ index compares the combined flux in the H and K emission lines with 
the flux in comparison bands at 3900 \AA\ and 4000 \AA. Such an index is 
not well suited to measurement from 
Hamilton echelle spectra since the four wavelength
regions used to obtain the index are all located in different echelle
orders. However, Figure~10, which shows a plot of both $K$ and $K_{\rm v}$ 
versus $S$ for the Mount Wilson survey stars from Table~1 
with discernible emission, indicates that
there is a relationship between the Hamilton-echelle indices and $S$, although 
this latter index does exhibit an
intrinsic spread among stars for which $K$ and $K_{\rm v}$ are close to unity. 
Although the errors in our values of $K$ and $K_{\rm v}$ are difficult to 
assess, since we have few measurements of stars observed on different nights,
Figure~10 does illustrate that the accuracy is sufficient to
distinguish between stars of high and low activity levels, which is the main 
purpose for which we wish to use these measurements. 

Comparisons between the \caII\ K-line emission strengths of the 
Mount Wilson stars from Table~1 and the subdwarfs from Table~2
are shown in Figure~11, in which both $S$ (upper panel) and $K_{\rm v}$ 
(lower panel) are plotted against $B-V$. 
In the upper panel, the one subdwarf shown (as an 
open circle) is HD~103095 (= Groombridge 1830); the point plotted is an
average of the largest
and smallest values of $S$ listed by Duncan et~al. (1991). 
Also shown in the upper panel are the mean locii for Population~I dwarfs
having weak \caII\ emission (solid line) 
and strong emission (dashed line). These locii
have been determined by eye from Figure~1 of Noyes et~al. (1984), in which
a bimodal distribution in $S$ is apparent among F and G dwarfs
in the Mount Wilson program. The 
level of emission in Groombridge 1830 is comparable to
that in relatively low-activity Population I dwarfs. 

In the lower panel of Figure~11
$K_{\rm v}$ is plotted versus $B-V$ for stars in Tables 1 (filled circles)
and 2 (open symbols) with discernible emission.
The two high-activity stars, HD 78366 and HD 101501, fall in a region of the
figure that is well separated from both the low-activity Population~I stars
and the subdwarfs.
The subdwarfs define a relatively tight sequence that overlaps the data for the
low-activity dwarfs sampled from the Mount Wilson survey. 
As would be expected on the basis of
their old age, the subdwarfs appear to be low-activity stars. Both
HD 153344 and HD 221613, which on the basis of their metallicities are 
probably disk dwarfs rather than subdwarfs, also lie along the low-activity
sequence in Figure~11.

In Figure~11 both the $S$ and $K_{\rm v}$ indices show a strong $B-V$ colour
dependence. The sequence defined by the subdwarfs in the lower panel
of Figure~11 rises with increasing $B-V$ colour. 
The $S$ index among low-activity Population~I dwarfs behaves in the same way
(see e.g., Noyes et al. 1984). In the case of the $S$ index this colour 
dependence is due to the reduction with decreasing effective temperature 
of the flux in the reference passbands (Middelkoop 1981).
The $K_{\rm v}$ index is presumably also responding in the same way; for a
given intensity $I_{\rm P1}$, the index will increase with increasing $B-V$
due to the systematic reduction in the comparison intensity $I_{\rm B1}$.

\section{Conclusions}

The main conclusion of this paper is that Population II subdwarfs do have
chromospheres, i.e., their outer atmospheres are heated by some form of
non-radiative process.
In terms of the contrast of the peaks in the K-line emission
profile relative to the neigbouring absorption profile,
and the preponderance of an asymmetry of $V/R > 1$, the subdwarfs appear 
similar to low-activity dwarfs sampled from the Mount Wilson survey. These 
observations
are consistent with data on the \mgII\ lines of metal-poor solar-type
stars reported by Peterson \& Schrijver (1997).

The origin of an asymmetry of $V/R > 1$ among both dwarfs and subdwarfs is not 
well understood. As Linsky (1980) has pointed out such an asymmetry can be 
produced by downward motions in the region of formation of the central K$_3$
absorption feature, or by upwards motions in the K$_2$ formation region if 
there are no systematic motions in the region of K$_3$ formation (see 
Ayres \& Linsky 1975 for the definitions of this terminology). Presumably
the reason for the $V/R >1$ asymmetry among the subdwarfs is the same as for
the Population~I dwarfs, including the Sun. However, the origin of this 
asymmetry even for the best studied case of the Sun is still not 
completely clear. This
subject has been reviewed by Linsky (1980), Cram (1983), and Rutten \&
Uitenbroek (1991). The concensus seems to be that the asymmetry is associated
with upwards propagating gravity or acoustic waves (see, e.g., Carlsson \&
Stein 1992, 1997). 

In addition to the uncertainty in the origin of the $V/R$ asymmetry,
the \caII\ K-line emission observations do not permit a determination of
whether the chromospheric activity
of the subdwarfs is being modulated by a magnetic dynamo, as for the Sun, or
whether it is produced by a process which is active even in the absence of 
such a dynamo. Statistical analyses of the chromospheric emission line 
fluxes of late-type Population~I stars have been used to argue that the emission
can be divided into two components
(Schrijver 1987a,b, 1995): a ``magnetic dynamo'' component which correlates with
stellar rotation rate and decreases with age, and a rotation-independent 
``basal'' component. The magnetic dynamo component 
is important for stars like the
Sun, and is particularly strong in late-type dwarfs younger than 1 Gyr.
The basal component is thought to reflect the contribution
of acoustic heating to the maintenance of a chromosphere (Schrijver 1987b;
Cuntz, Rammacher, \& Ulmschneider 1994). 

Whether the chromospheres of subdwarfs are ``basal chromospheres'' or are 
maintained by a magnetic dynamo is not clear from the \caII\ K line 
observations. The finding by Baliunas et~al. 
(1995) that the subdwarf HD~103095 exhibits an
activity cycle of period 7.3 years in the \caII\ H and K emission lines, 
suggests an analogy with the solar cycle, implying
that the subdwarfs have chromospheres whose long-term
behaviour is dominanted by a magnetic dynamo.

The similarity in \caII\ K-line emission strength among the subdwarfs and the
lowest-activity stars sampled from the Mount Wilson survey seems consistent 
with the spectroscopy of main-sequence stars in the old open
clusters NGC 188 and M67 by Barry, Cromwell, \& Hegge (1984) which suggests
that the rate of decline of the fluxes in the Ca II H and K line cores 
flattens out after $\sim$ 3 Gyr. These observations, and those noted in the
previous paragraph, would be consistent with a scenario in which the magnetic 
dynamo ceases to decline significantly in activity once main-sequence stars 
reach an age of around 5 Gyr. There could be two reasons for this. 
Either the dynamo reaches a true minimum level, as in the turbulent magnetic 
field scenario of Durney, De Young, \& Roxburgh (1993), or else the rate of 
spin down of dwarfs and subdwarfs older than 5 Gyr becomes very slow. 

The alternative to these dynamo scenario is that chromospheres among stars 
older than
5 Gyr are largely maintained by an age-invariant basal heating mechanism, and 
that activity produced by an age-dependent magnetic dynamo has dropped to a 
low level by comparison. In sufficiently old main-sequence subdwarf stars the 
dynamo might be relatively inactive and acoustic heating may become the dominant
mechanism for maintaining a ``basal chromosphere.'' 
This is the scenario proposed by Peterson \& Schrijver
(1997) on the basis of similarities observed between the \mgII\ emission line 
profiles of metal-poor solar-type stars and solar quiet regions. In this respect
it is perhaps relevant to note that the most extreme $V/R >1$ \caII\
K line asymmetry in
the integrated spectrum of the solar disk occurs around the times of
solar minimum (White \& Livingston 1981).
 
The type of observation that may differentiate between these scenarios
is a search for soft x-ray emission among subdwarfs. If subdwarfs are found to 
have solar-like
coronae, with soft x-ray properties consistent with temperatures of 
1-2$\times 10^6$ K, then the case for dynamo-maintained chromospheres and 
coronae would be strengthened. 

We thank the referee for a number of useful comments and for drawing our 
attention to a number of relevant references in the literature.

\onecolumn

\begin{table}[ht]
\centerline{Table 1: K Emission Line Properties of Mount Wilson Survey Stars}
\begin{center}
\begin{tabular}{lllllcccl}\\
\hline\hline\noalign{\smallskip}
\multicolumn{1}{c}{Name}    & 
\multicolumn{1}{c}{$V$}     &   
\multicolumn{1}{c}{$B-V$}   &  
\multicolumn{1}{c}{SpT}     &    
\multicolumn{1}{c}{$S$}     &    
\multicolumn{1}{c}{Emission} &  
\multicolumn{1}{c}{$V/R$}   &
\multicolumn{1}{c}{$K$}     &
\multicolumn{1}{c}{$K_{\rm v}$} \\
\noalign{\smallskip}\hline\noalign{\medskip}                           
HD 13421   &  5.63  &   0.56  &  G0IV   & 0.130  &   yes$^f$  &   $>1$   &
   ...     &  $\leq 1^g$  \\   
HD 78366   &  5.93  &   0.60  & F9V     & 0.240  &   yes$^d$  &    $>1$  &
  1.99     &  2.02  \\  
HD 101501  &  5.33  &   0.72  & G8V     & 0.310  &   yes$^d$  &    $>1$  &  
  2.60     &  2.74  \\  
HD 115617  &  4.74  &   0.71  & G6V     & 0.160  &   yes      &    $>1$  & 
  1.13     &  1.16  \\
HD 136202  &  5.06  &   0.54  & F8III-IV & 0.140 & infl$^b$   & $\approx 1$  &
  ...      &  $\leq 1^g$  \\
HD 141004  &  4.43  &   0.60  & G0V     & 0.160  &   yes      &    $>1$   &
  1.09     &  1.17  \\
HD 142373  &  4.62  &   0.56  & F8V     & 0.142  &   yes      &    $>1$   &
  ...      &  1.06  \\
HD 159332  &  5.64  &   0.48  & F6V     & 0.145  & infl$^a$   &    $>1$   &
  ...      &  $\leq 1^g$   \\
HD 176051  &  5.22  &   0.59  & F9V     & 0.180  &   yes      &    $<1$   & 
  1.12     &  1.09  \\
HD 184499  & 6.63$^e$ & 0.60$^e$  &  G0V$^e$  & 0.145$^e$ & Kn,Hy$^c$ & $>1$  &
  ...      &  $\leq 1^g$   \\
HD 187013  &  4.99  &   0.47  & F7V     & 0.150  &  infl$^a$  &    $>1$   &
  ...      &  $\leq 1^g$   \\   
HD 187691  &  5.11  &   0.55  & F8V     & 0.150  & Kn,Hy$^c$  &    $>1$   &
  ...      &  $\leq 1^g$   \\
HD 190406  &  5.80  &   0.61  & G1V     & 0.190  &   yes      &    $\geq 1$   &
  1.17     &  1.19  \\
HD 207978  &  5.53  &   0.42  & F6IV-V  & 0.155  &    no      &    ...    &
  ...      &  ...   \\
\noalign{\medskip}\hline
\end{tabular}
\end{center}

\begin{center}
\begin{minipage}{100mm}
\small$^a$ K$_{\rm 2v}$ profile shows an inflection near the core but no 
      emission peak having a local maximum.\\
\small$^b$ Both K and H profiles show an inflection near the core (although the
      inflection in the H line is less obvious), but no emission peaks.\\
\small$^c$ Emission in H line, but K line shows only an inflection in the 
      core.\\
\small$^d$ Very strong emission.\\ 
\small$^e$ Photometry and spectral type from Hipparcos Input Catalogue. The 
      value of $S$ is an average of the largest and smallest 
      values tabulated by Duncan et al. (1991).\\
\small$^f$ Weak emission in violet wing (K$_{\rm 2v}$) of the K$_2$ line.
      Possible H$_{\rm 2v}$ emission, but very weak.\\
\small$^g$ K line profile shows either no emission or an inflection near the
      line core without a clear emission peak. 
\end{minipage}
\end{center}
\end{table}

\begin{table}[ht]
\centerline{Table 2: K Emission Line Properties of Metal-Poor Dwarfs}
\begin{center}
\begin{tabular}{llclcclll}\\
\hline\hline\noalign{\smallskip}
\multicolumn{1}{c}{Name}    & 
\multicolumn{1}{c}{G}       &
\multicolumn{1}{c}{$V$}     &   
\multicolumn{1}{c}{$B-V$}   &  
\multicolumn{1}{c}{[Fe/H]}     &    
\multicolumn{1}{c}{Emission} &  
\multicolumn{1}{c}{$V/R$}   &
\multicolumn{1}{c}{$K$}     &
\multicolumn{1}{c}{$K_{\rm v}$} \\
\noalign{\smallskip}\hline\noalign{\medskip}                           
HD 4906  &  G32-53  &  8.76  &  0.76  &  $-1.1$  &  yes  &  $\geq 1$ &
    1.26 & 1.31 \\
BD $+29$ 366 &  G74-5 & 8.76 &  0.58  &  $-1.1$  & infl$^a$ & $> 1$  &
    ... & $\leq 1^i$ \\   
BD $-1$ 306 & ... & 9.08 & 0.58 & $-1.0$ & no & ... & ... & $\leq 1^i$ \\ 
HD 14056 & G4-10  & 9.05 & 0.62 & $-0.8$ & infl$^b$ & $> 1$ & ... & $\leq 1^i$ \\
HD 20039 & G221-7 & 8.89 & 0.75 & $-1.2$ & yes & $> 1^c$: & ... & 1.35 \\
HD 20507 & G245-73 & 6.93 & 0.45 & $-1.0$ & no & ... & ... & $\leq 1^i$\\
HD 25329 & ... & 8.51 & 0.87 & $-1.5$ & yes & $\approx 1^c$: & ... & ... \\
BD $+37$ 1458 & ... & 8.92 & 0.60 & $-2.4$ & no & ... & ... & $\leq 1^i$\\
HD 64606 & G112-54 & 7.43 & 0.73 & $-0.9$ & yes$^d$: & $> 1$ & 1.22 & 1.25 \\
HD 78050 & G9-47 & 7.68 & 0.80 & $-1.9$ & yes & $> 1$ & 1.52 & 1.60 \\
HD 103095 & ... & 6.44 & 0.75 & $-1.2$ & yes & $> 1$ & ... & ... \\
HD 103912 & G122-57 & 8.36 & 0.86 & $-1.7$ & yes & $\geq 1^e$ & 2.04$^h$ &
   1.86$^h$ \\
HD 134169 & ... & 7.68 & 0.55 & $-1.6$ & yes$^d$: & $> 1$ & ... & 1.06 \\
HD 153344 & G226-36 & 7.07 & 0.67 & $-0.3$ & yes & $> 1$ & 1.05 & 1.08 \\
HD 157948 & G182-7  & 8.10 & 0.76 & $-1.3$ & yes & $> 1$ & ... & ... \\
BD $+23$ 3130 & G170-47 & 8.95 & 0.64 & $-2.9$ & no & ... & ... & $\leq 1^i$ \\
HD 188510 & ... & 8.83 & 0.59 & $-1.6$ & no & ... & ... & $\leq 1^i$ \\
HD 194598 & ... & 8.35 & 0.49 & $-1.3$ & no & ... & ... & $\leq 1^i$ \\
HD 198300 & G230-49 & 8.51 & 0.59 & $-0.9$ & yes$^d$: & ... & ... & ... \\
HD 200580 & G25-15 & 7.34 & 0.54 & $-1.0$ & no & ... & ... & $\leq 1^i$ \\
HD 201891 & ... & 7.37 & 0.51 & $-1.1$ & no & ... & ... & $\leq 1^i$ \\
HD 221613 & G171-3 & 7.14 & 0.58 & $-0.6$ & Ky,Hn$^f$ & $< 1^f$ & 1.05 & 1.01 \\
HD 224087 & G129-42 & 8.94 & 0.81 & $-0.7$ & yes & $> 1^g$ & ... & ...\\

\noalign{\medskip}\hline
\end{tabular}
\end{center}

\begin{center}
\begin{minipage}{100mm}
\small$^a$ Both K$_2$ and H$_2$ profiles show an inflection near the core but 
      no emission peak having a local maximum.\\
\small$^b$ Inflection in violet side of the core of the K line.\\
\small$^c$ Noisy spectrum.\\
\small$^d$ Emission is very weak.\\
\small$^e$ Two spectra were obtained; in April 1995 the spectra show a nearly
           symmetric $K_2$ emission feature with the violet wing having only 
           a slightly-higher peak than the red wing (Figure 5d), while in June 
           1995 a clearer asymmetry of $V/R > 1$ was exhibited.\\
\small$^f$ Emission in core of the K line, but not the H line. The quoted
           $V/R$ refers to the K line.\\
\small$^g$ Emission profile shows an asymmetric emission peak, but does 
           not exhibit a central K$_3$ absorption feature.\\
\small$^h$ Values of $K$ and $K_{\rm v}$ derived from the April 1995 spectrum.\\
\small$^i$ Inflection or no emission in core.\\ 

\end{minipage}
\end{center}
\end{table}

\begin{figure}[th]
\plotone{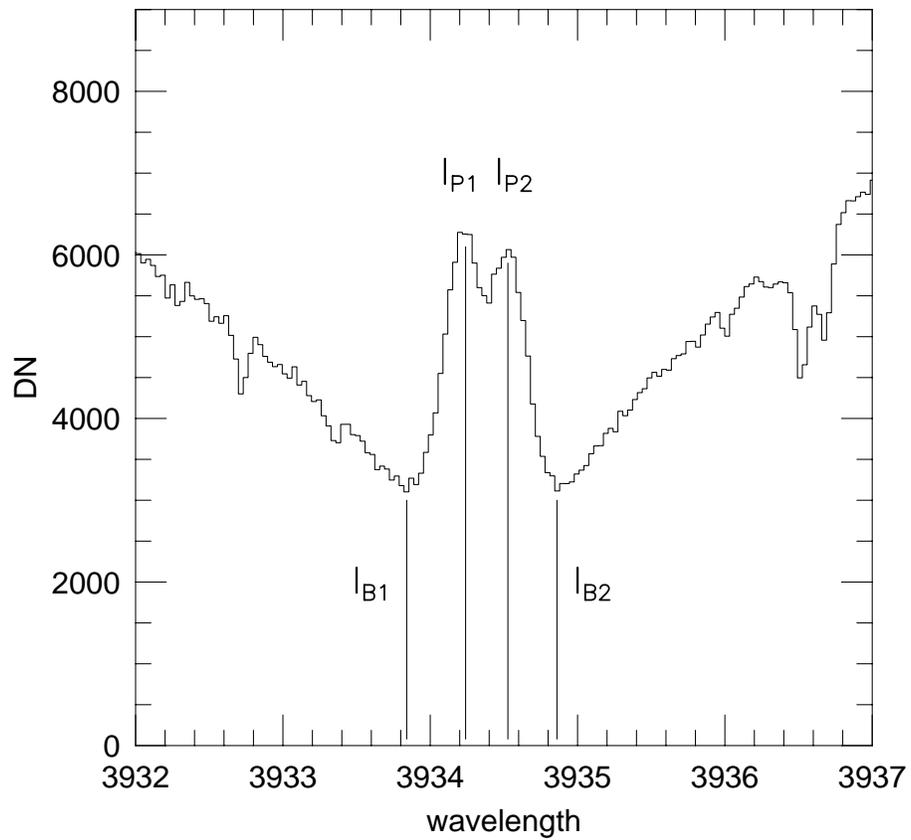}
\vglue -1.0in
\caption{The parameters used to calculate the emission-profile indices
$K$ and $K_{\rm v}$ are defined in this figure. The spectrum shown is
that of HD~78366. The wavelength scale corresponds to the rest frame
of the ThAr lamp used in wavelength calibrating the spectrum.}
\end{figure}

\begin{figure}[th]
\plottwo{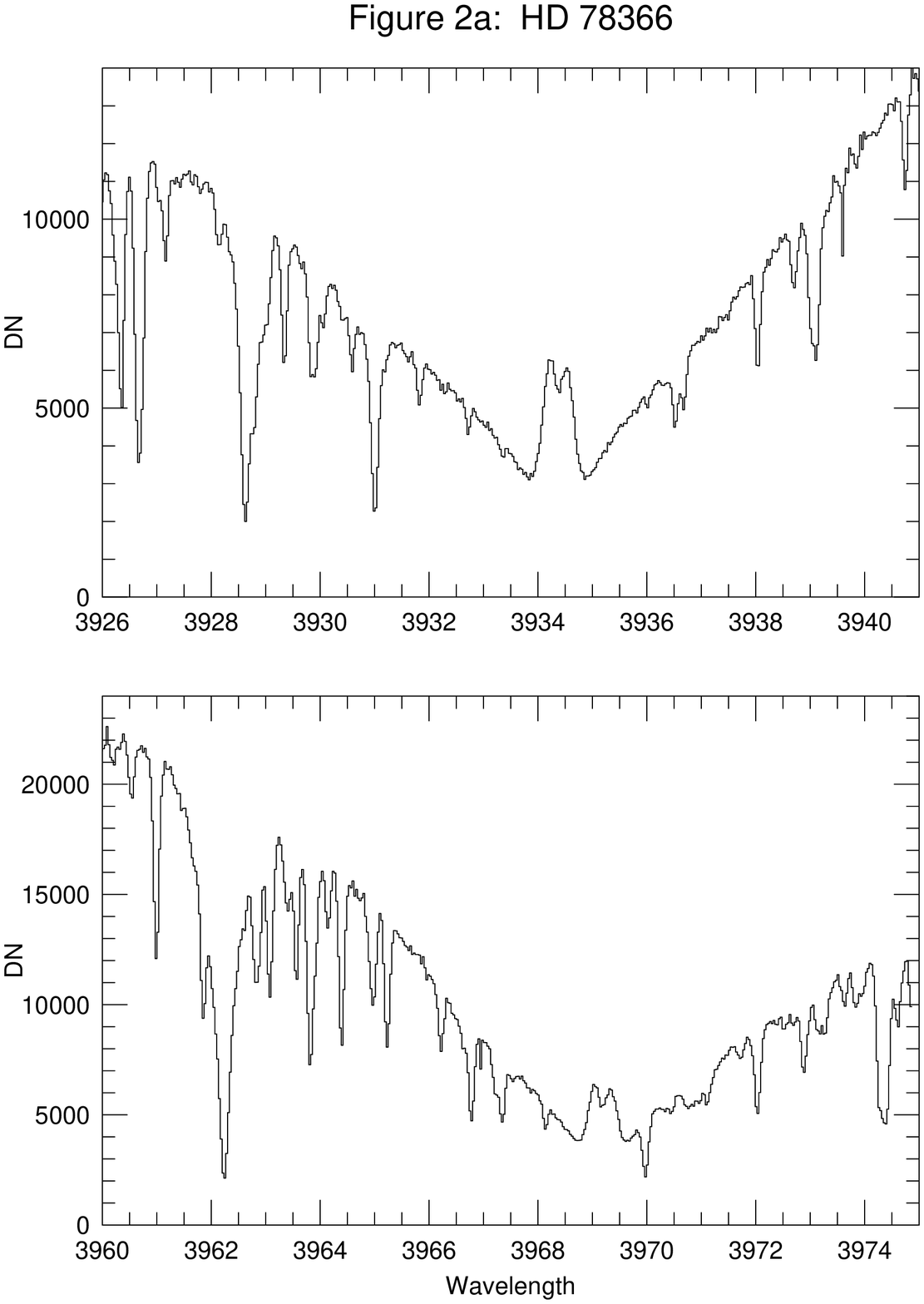}{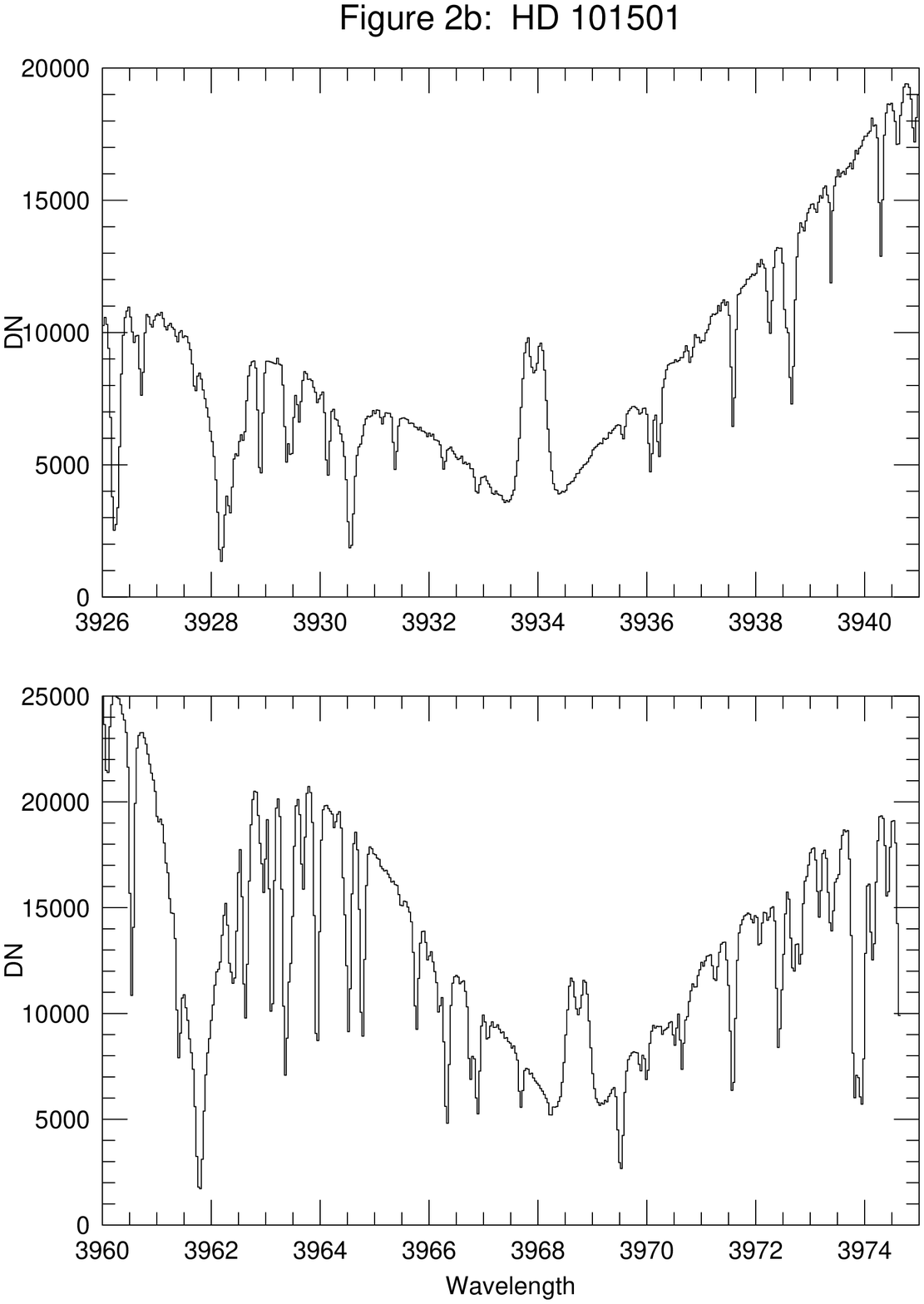}
\caption{Spectra of the \caII\ H and K lines for the dwarfs HD 78366
(panel a) and HD 101501 (panel b). These stars shows very strong H and
K emission lines as well as the largest values of $S$, $K$, $K_{\rm
v}$ among the stars in Table~1.}
\end{figure}

\begin{figure}[th]
\plottwo{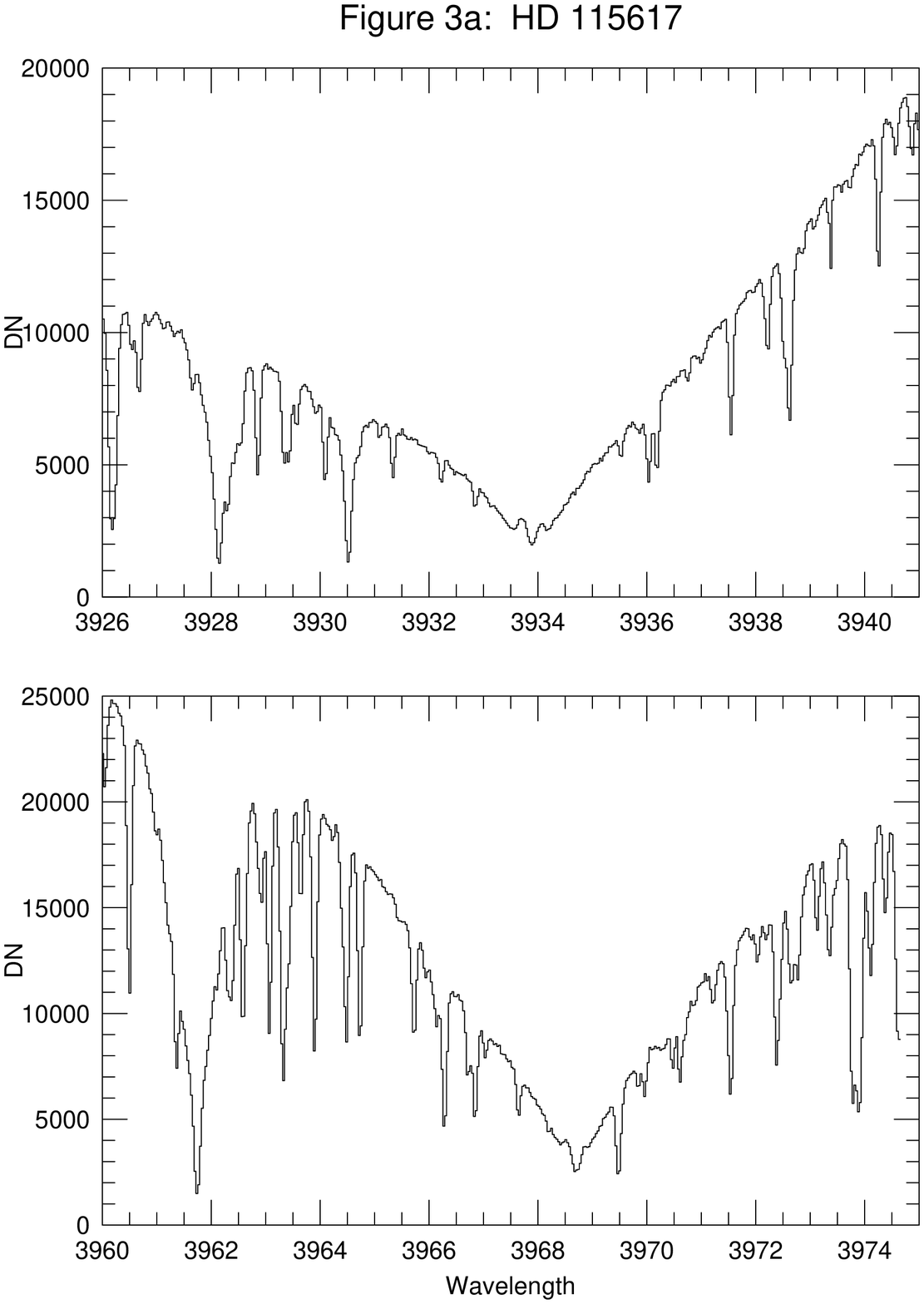}{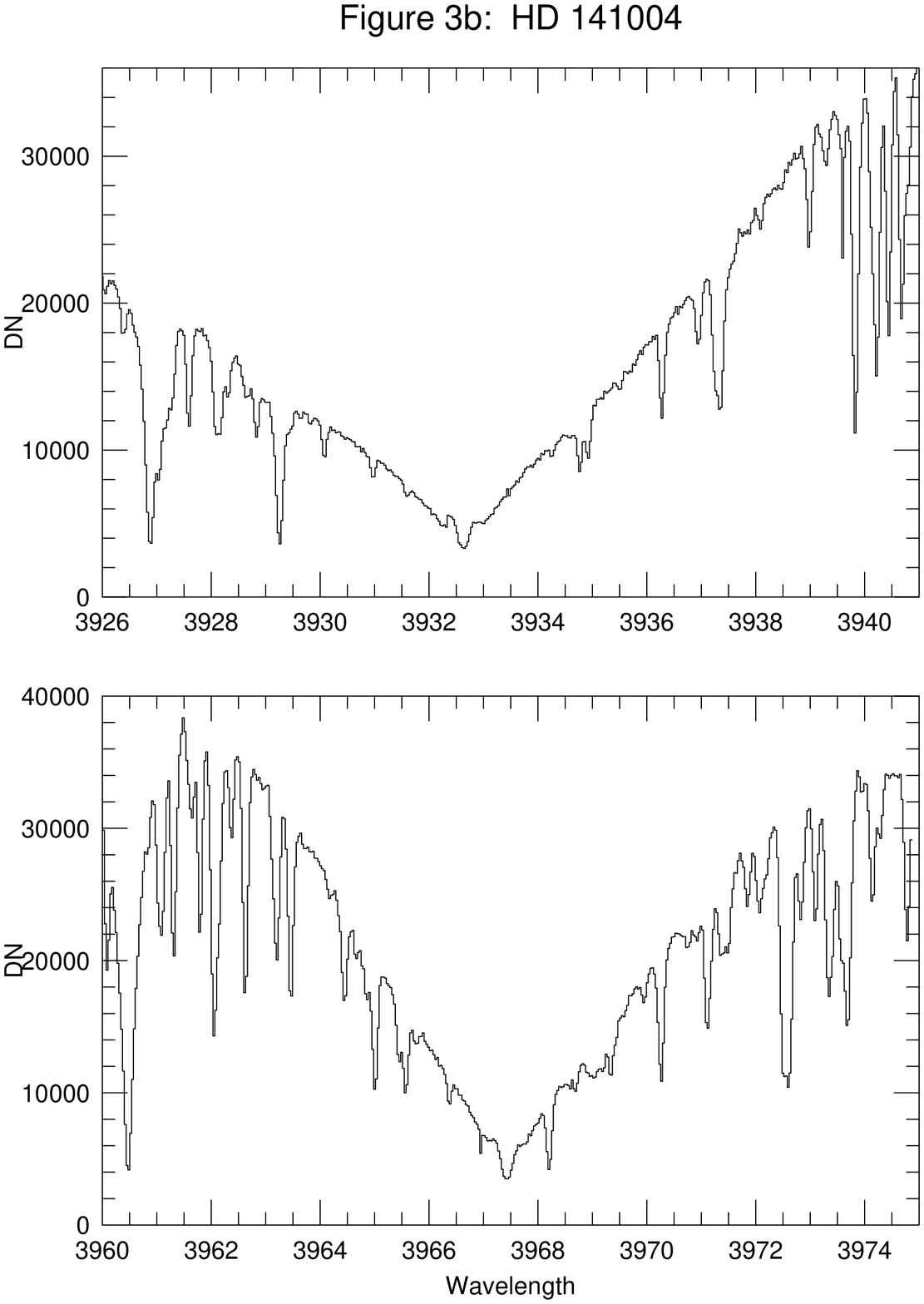}
\caption{Spectra of the \caII\ H and K lines for two low-activity
dwarfs from the Mount Wilson survey: HD 115617 (panel a) and HD 141004
(panel b).}
\end{figure}

\begin{figure}[th]
\plottwo{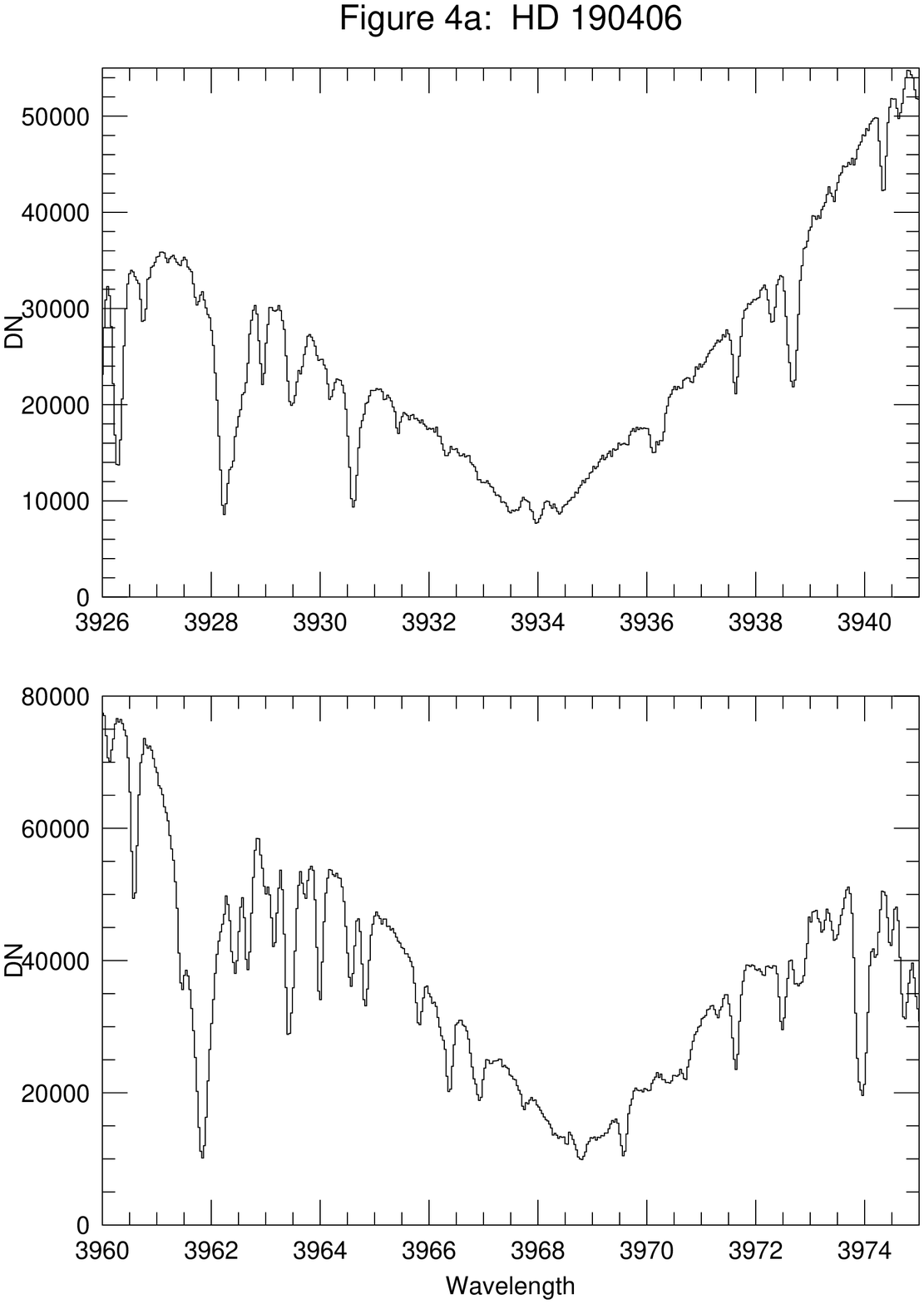}{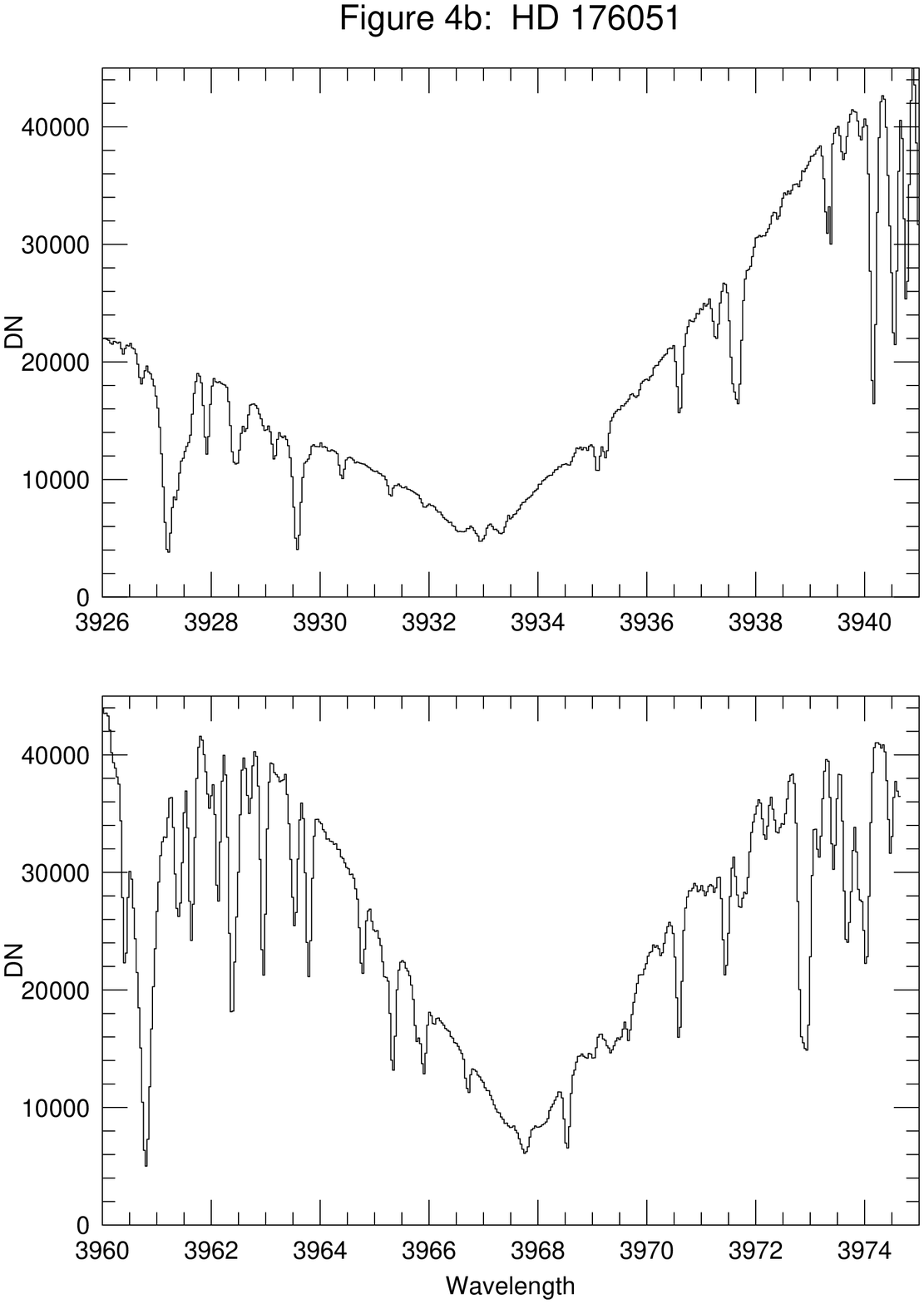}
\caption{Spectra of the \caII\ H and K lines for the dwarfs HD 190406
(panel a), for which $V/R$ is only slightly greater than unity (as
illustrated best by the H$_2$ line), and HD 176051 (panel b), which
unlike the other stars from Table~1 exhibits an asymmetry of $V/R <
1$.}
\end{figure}

\begin{figure}[th]
\plottwo{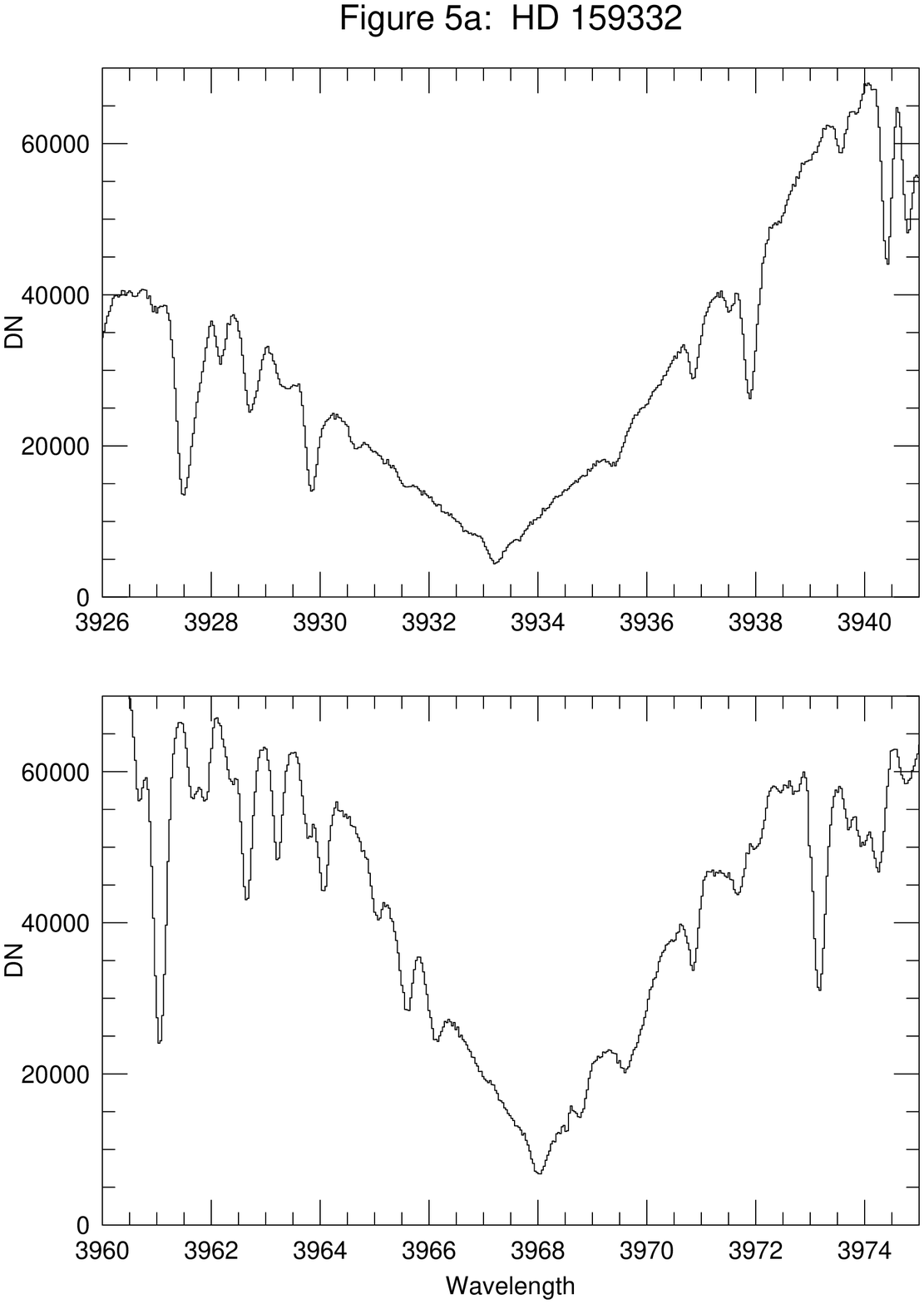}{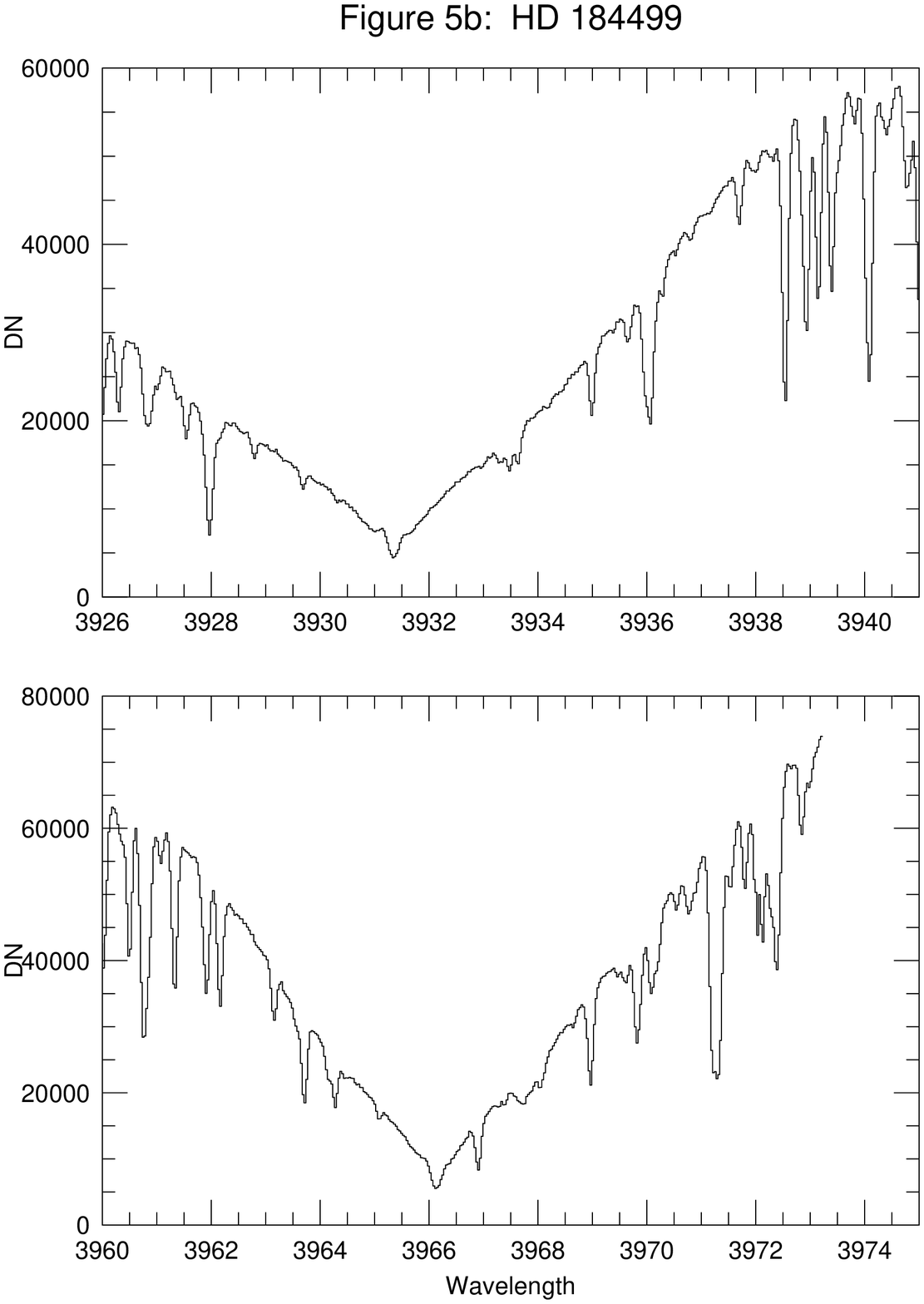}
\caption{Spectra of two dwarfs from the Mount Wilson survey that
exhibit inflections in the slope of the H and/or K lines near the line
center.  Panel (a) shows a case of inflections in both line cores,
whereas HD~184499 exhibits this phenomenon in the K line only.}
\end{figure}

\begin{figure}[th]
\plottwo{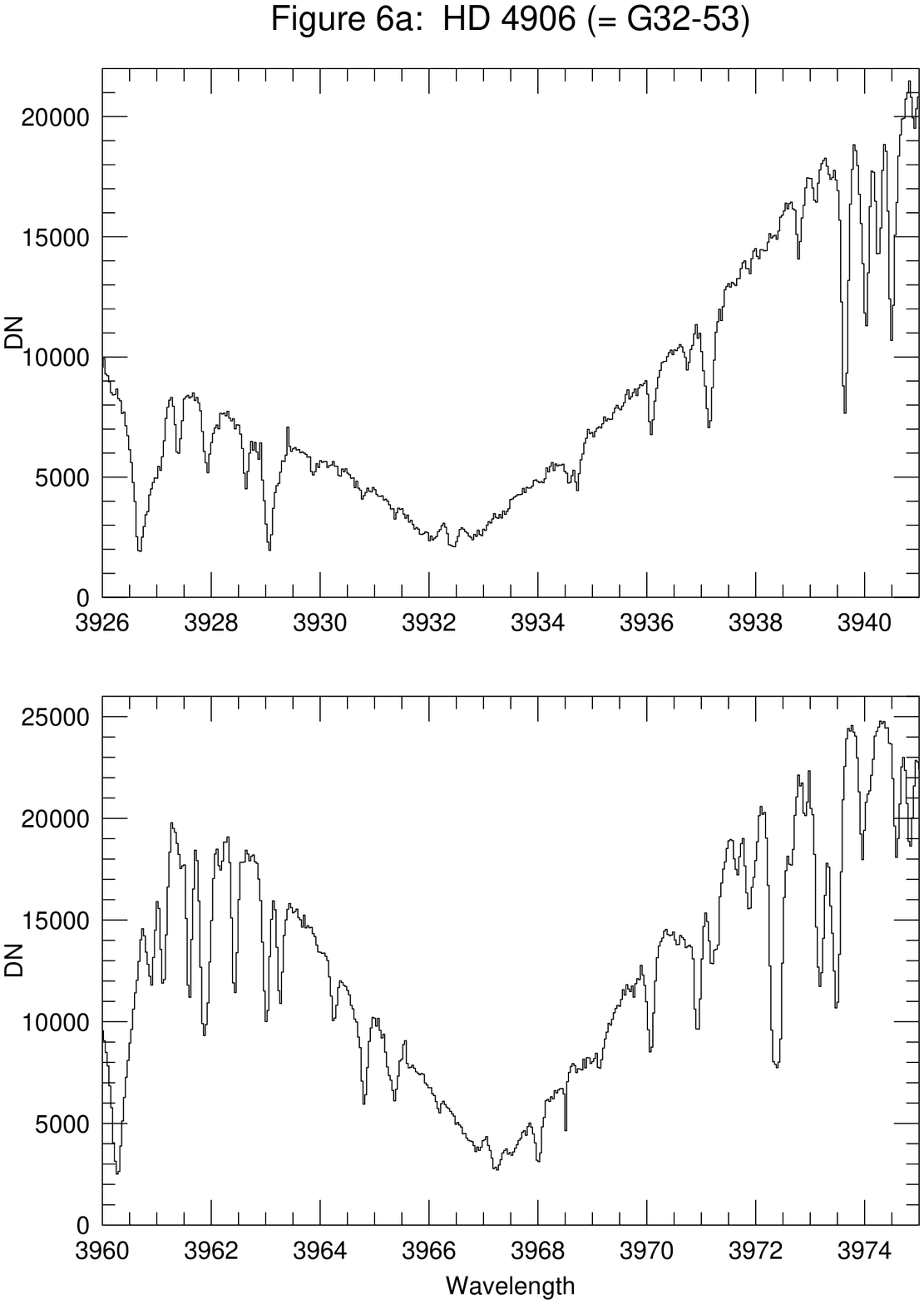}{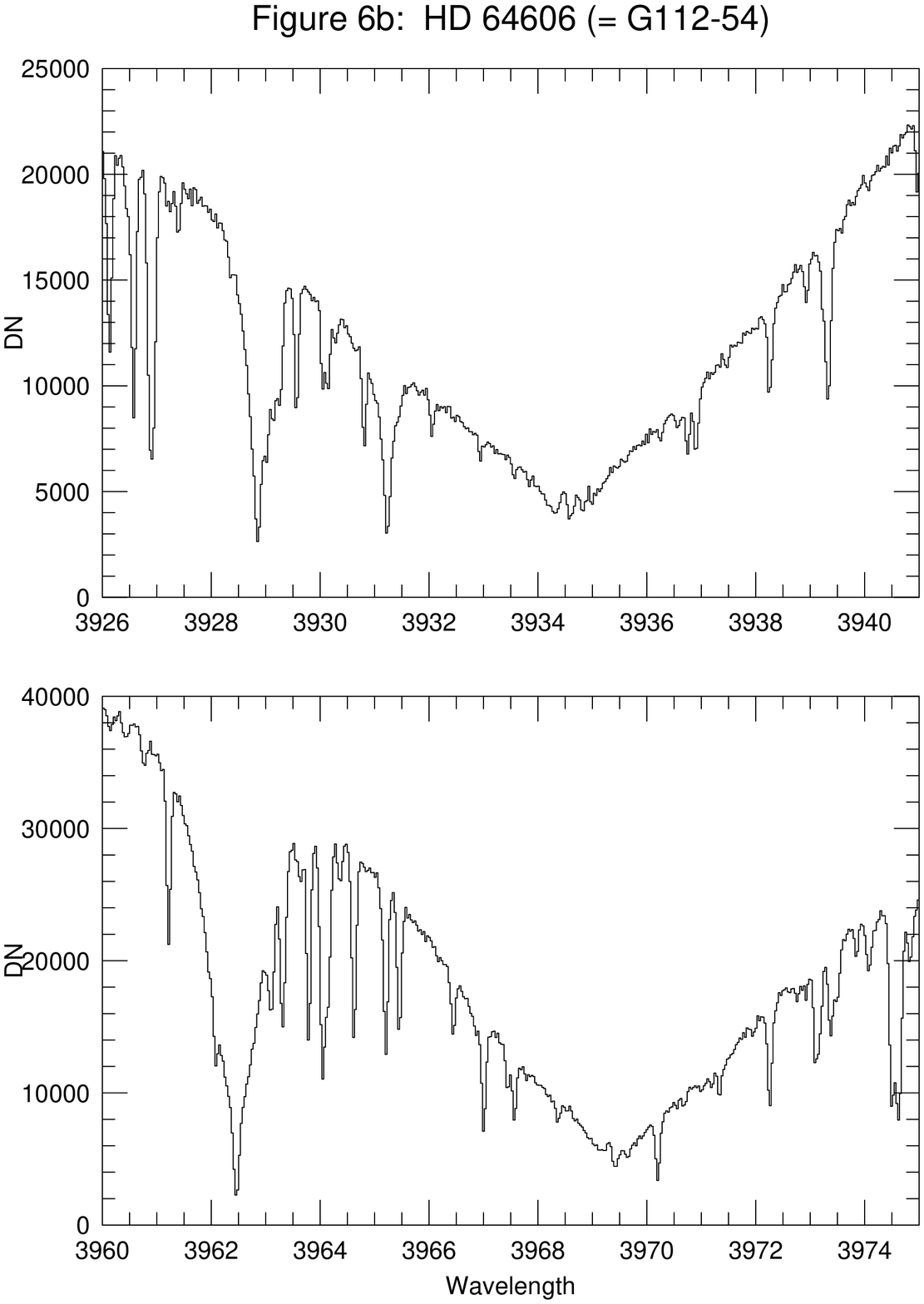}
\end{figure}
\begin{figure}
\vglue -1.2in
\plottwo{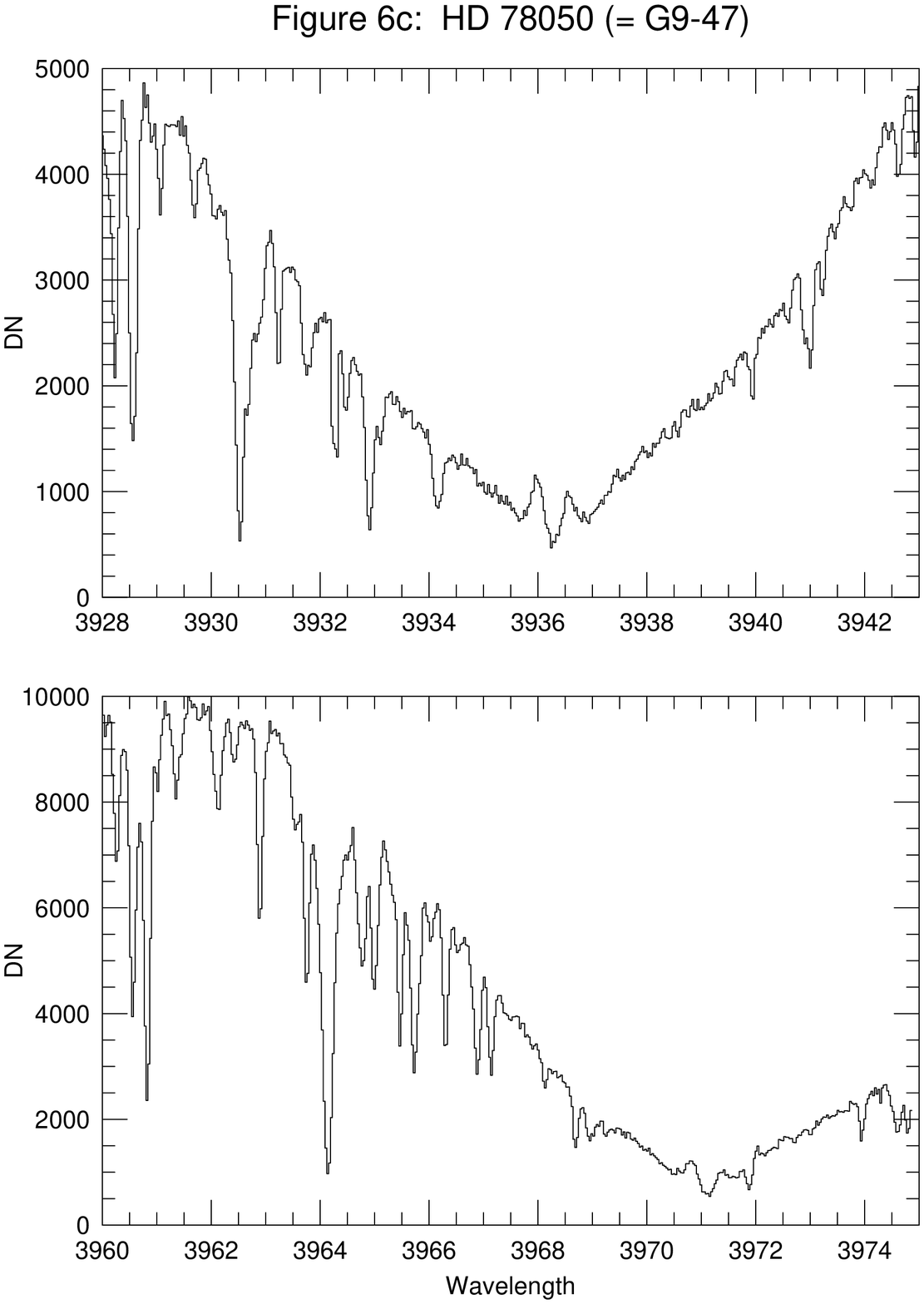}{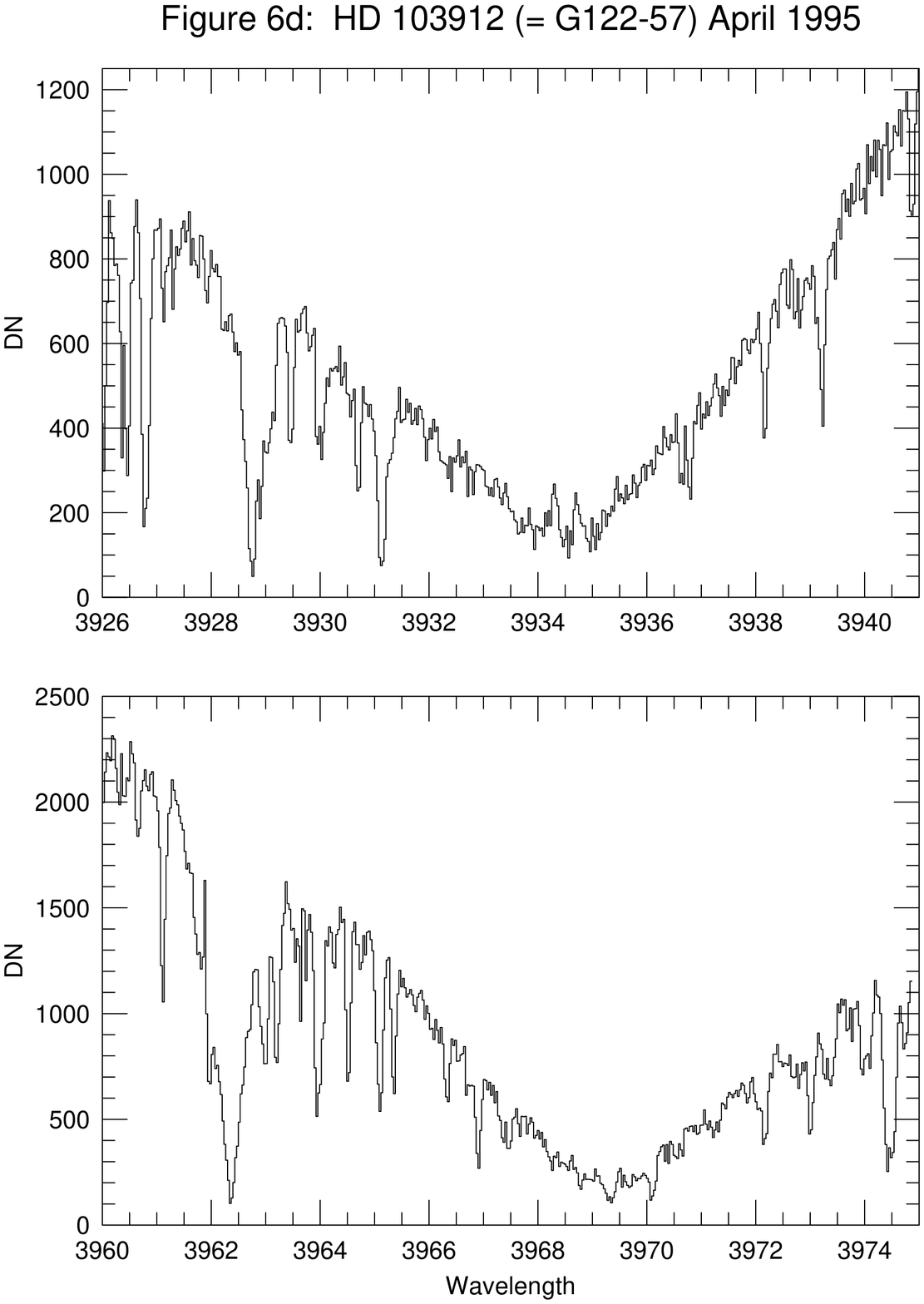}
\vglue -0.4in
\caption{Spectra of the \caII\ H and K lines for the subdwarfs HD 4906
(G32-53; panel a), HD 64606 (G112-54; panel b), HD 78050 (G9-47; panel
c), and HD 103912 (G122-57; panel d). These stars have colours of $B-V
\geq 0.70$, and all show emission in at least one of the line cores.}
\end{figure}

\begin{figure}[th]
\plottwo{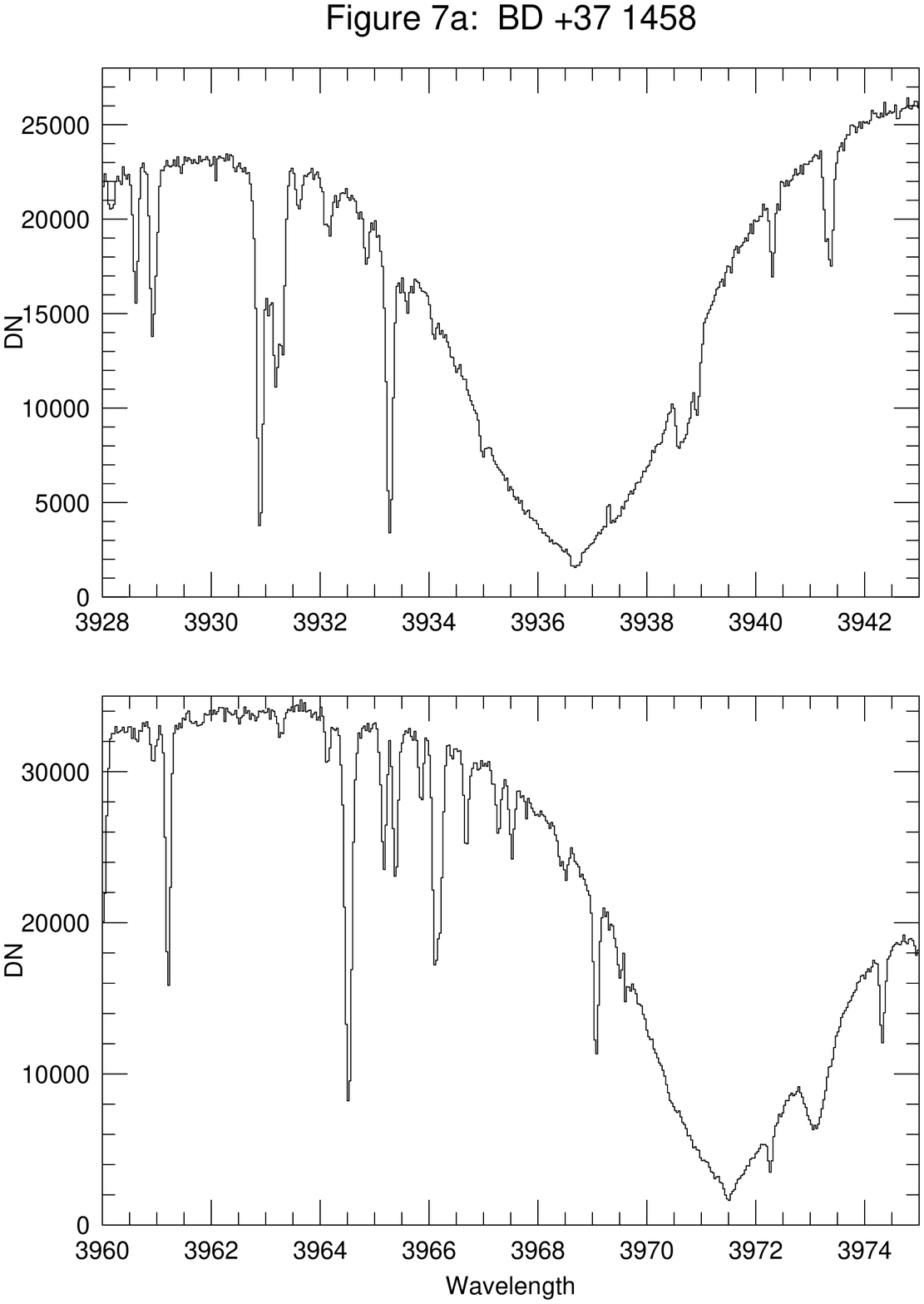}{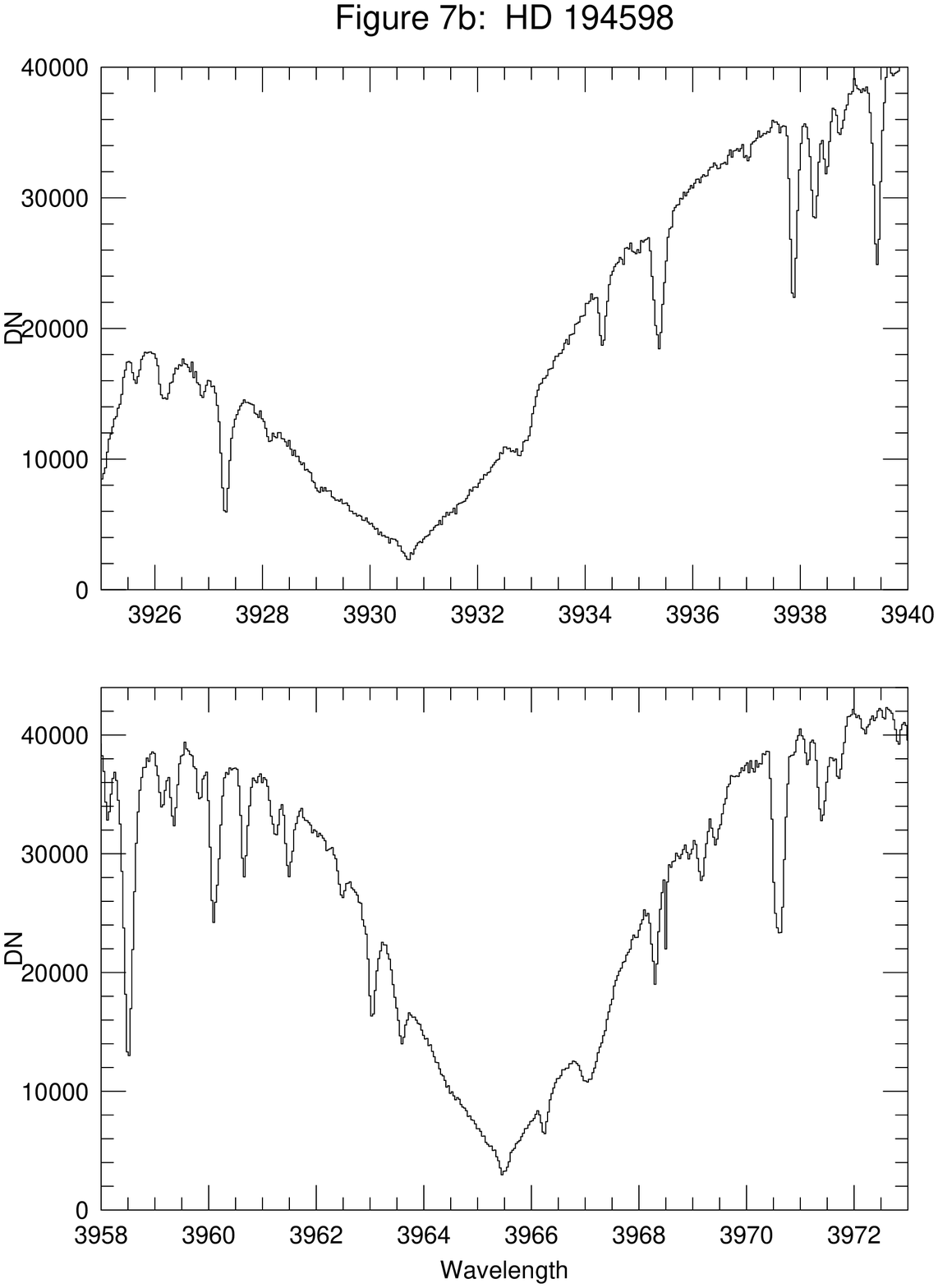}
\end{figure}
\begin{figure}
\vglue -1.2in
\plottwo{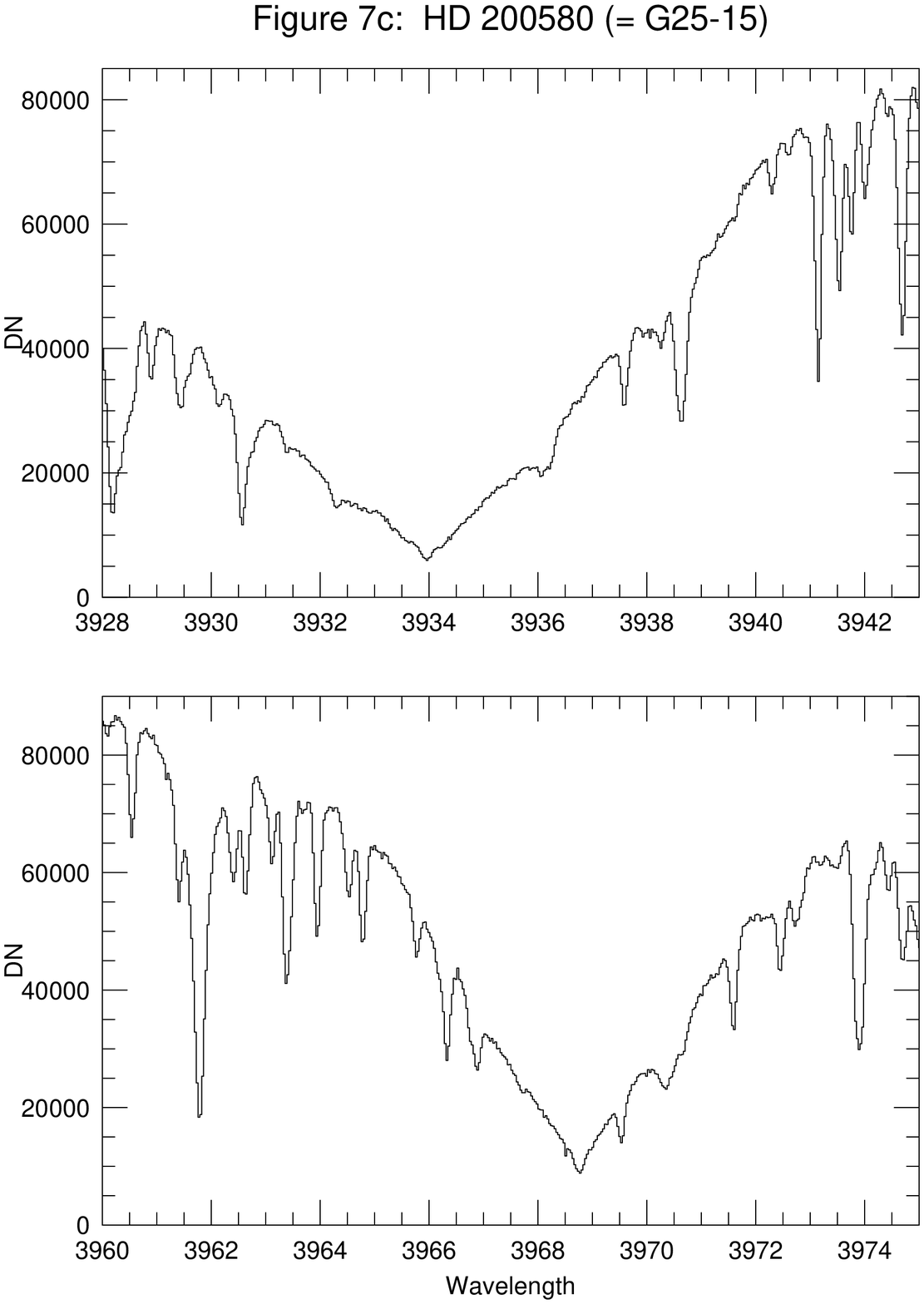}{fig7d.ps}
\vglue -0.4in
\caption{Spectra of the \caII\ H and K lines for three subdwarfs which
do not show emission in the line cores: BD $+37$ 1458 (panel a), HD
194598 (panel b), and HD 200580 [= G25-15] (panel c).}
\end{figure}

\begin{figure}[th]
\plottwo{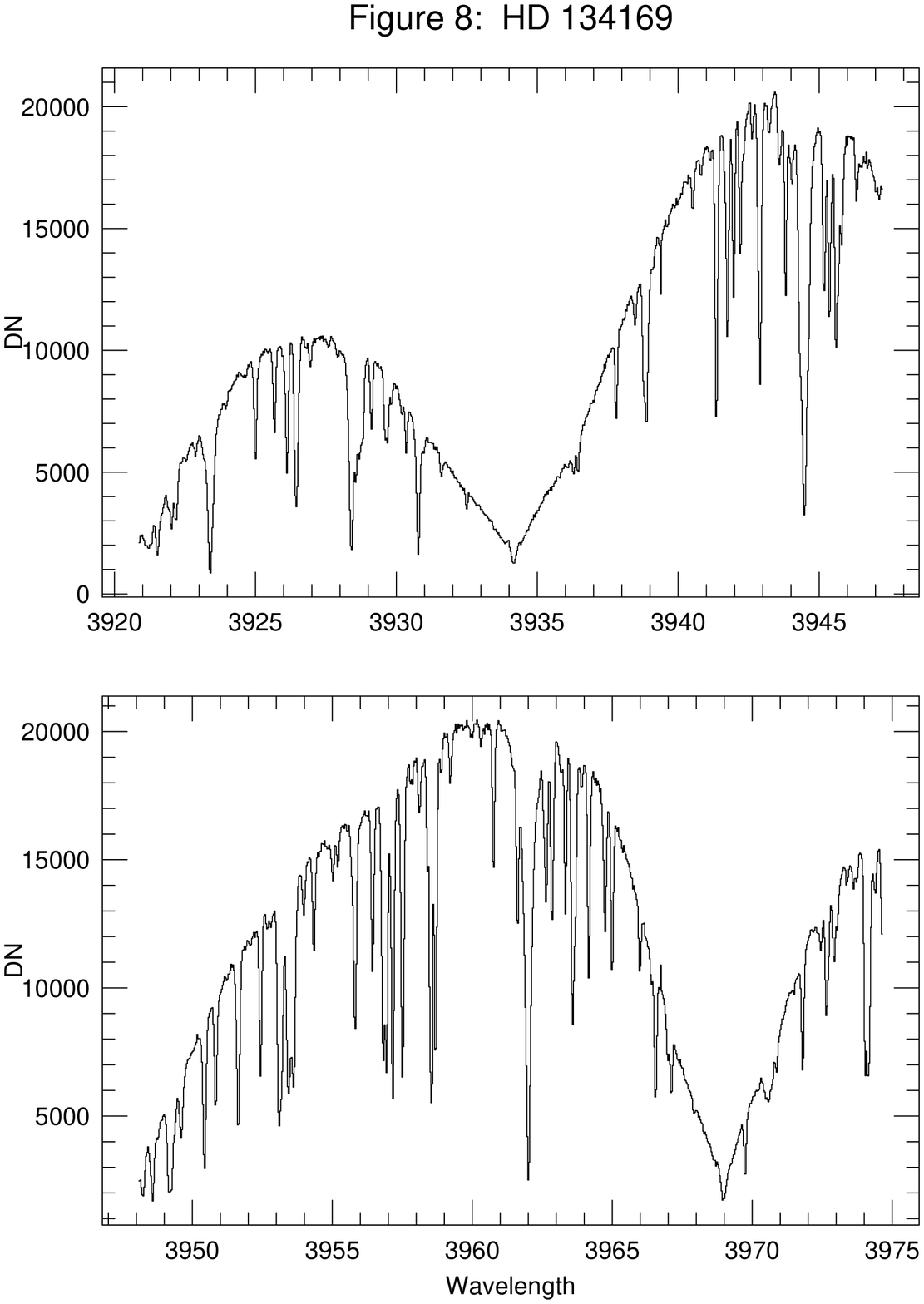}{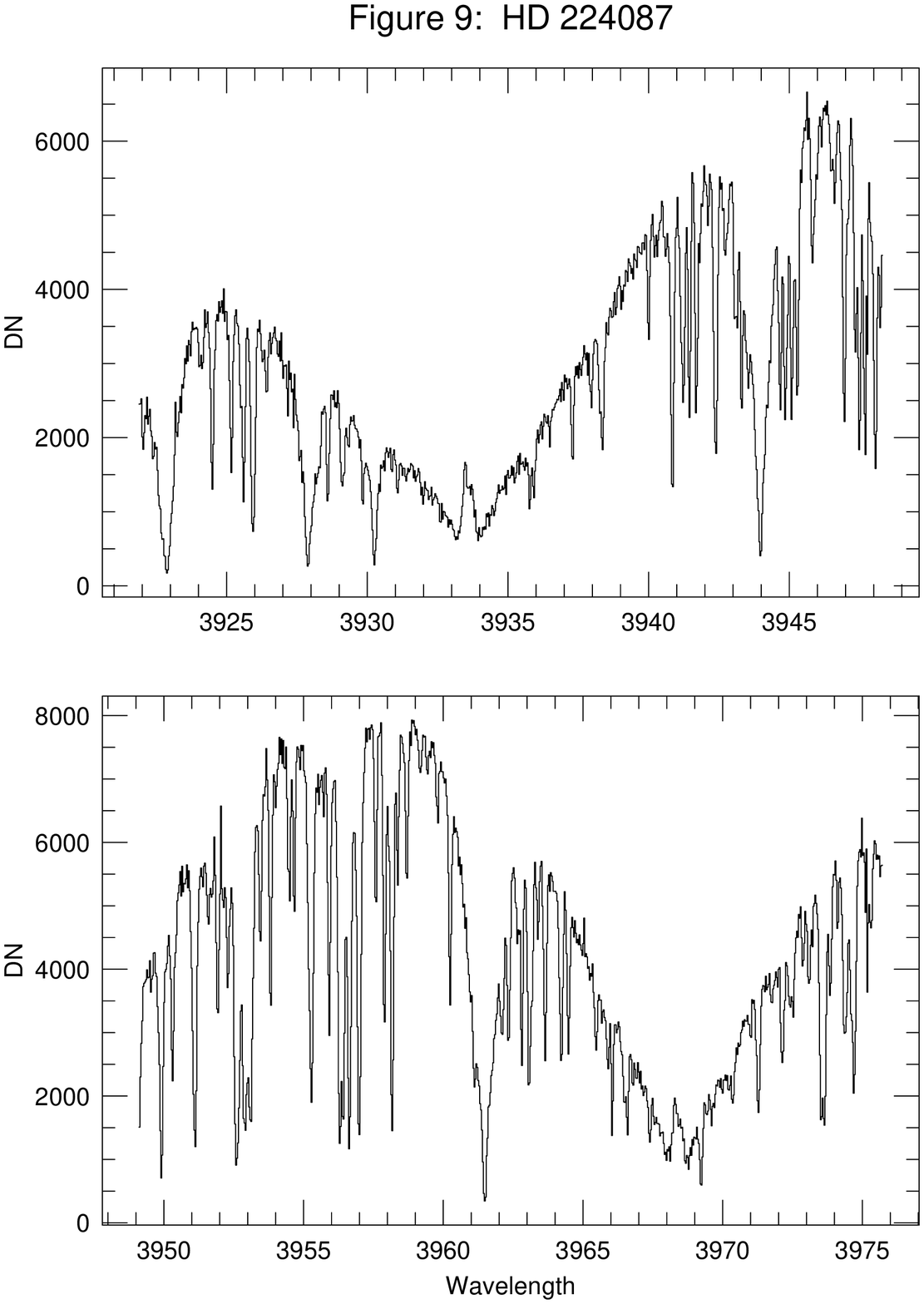}
\caption{Spectra of the \caII\ H and K lines of the subdwarf
HD~134169.}
\caption{Spectra of the \caII\ H and K lines of HD~224087.}
\end{figure}

\begin{figure}[th]
\plotone{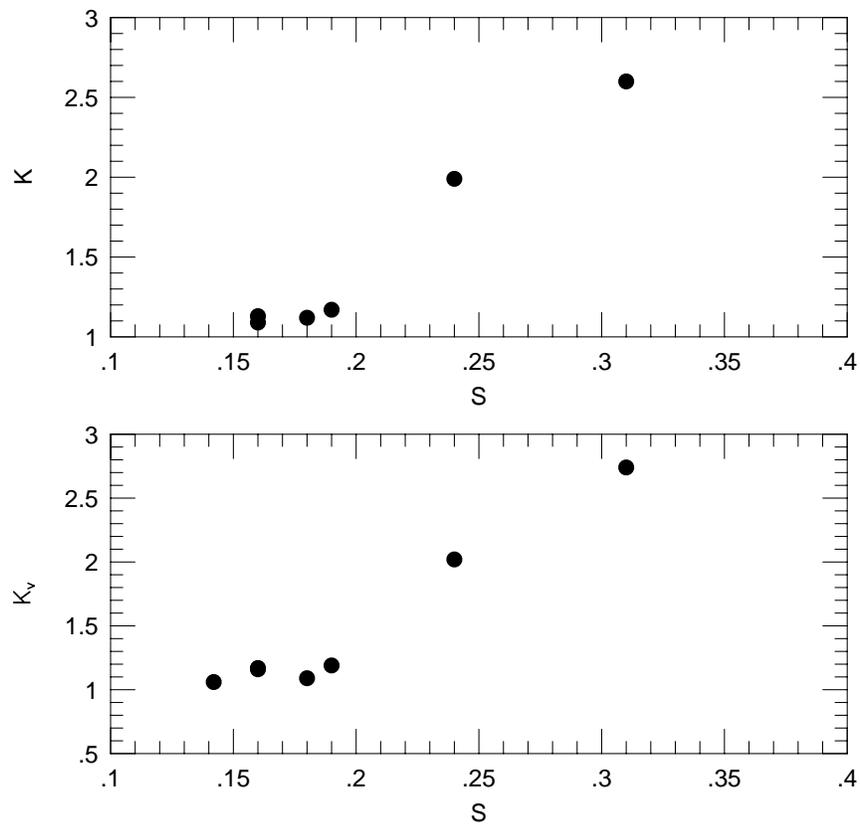}
\vglue -1.0in
\caption{The two echelle-based \caII\ K-line parameters $K$ and $K_{\rm
v}$ are shown plotted versus the Mount Wilson $S$ index for stars from
Table~1. Only stars with $K_{\rm v} > 1$ are shown; these stars have
distinct emission maxima in the profile of the K-line core.}
\end{figure}

\begin{figure}[th]
\plotone{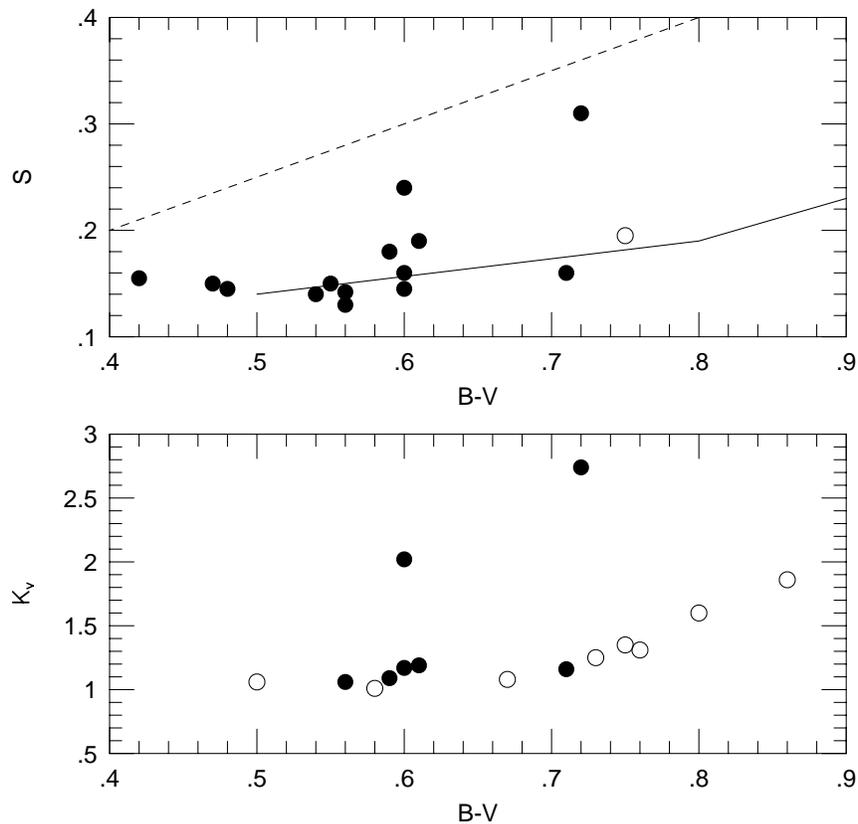}
\vglue -1.3in
\caption{(Upper panel.) The $S$ index for stars from Table~1 (filled
circles) and HD~103095 (open circle) is shown plotted versus
$B-V$. This figure indicates that the subdwarf HD~103095 (Groombridge
1830) has \caII\ H and K emission lines comparable in strength to
those of low-activity Population~I dwarfs. The solid and dashed lines
represent eye estimates of the mean locii for the low-activity and
high-activity dwarfs respectively from Noyes et~al. (1984; see their
Figure~1).  (Lower panel.) The K$_{\rm v}$ parameter is plotted versus
$B-V$ colour for Mount-Wilson-survey dwarfs from Table~1 (filled
circles) and subdwarfs from Table ~2 (open symbols).  Only stars with
$K_{\rm v} > 1$ are plotted.}
\end{figure}
 

\begin{thebibliography}{}

\vspace{3mm}
\parskip 0pt
\bibitem[dummy]{dummy01} Ayres, T. R., Linsky, J. L., 1975, ApJ, 200, 660
\bibitem[dummy]{dummy02} Baliunas, S. L., et al., 1995, ApJ, 438, 269
\bibitem[dummy]{dummy03} Barry, D. C., Cromwell, R. H., Hegge, R. H., 1984, ApJ, 277, L65
\bibitem[dummy]{dummy04} Basri, G. S., Linsky, J. L., 1979, ApJ, 234, 1023
\bibitem[dummy]{dummy05} Carlsson, M., Stein, R. F., 1992, ApJ, 397, L59
\bibitem[dummy]{dummy06} Carlsson, M., Stein, R. F., 1997, ApJ, 481, 500 
\bibitem[dummy]{dummy07} Carney, B. W., 1979, ApJ, 233, 211
\bibitem[dummy]{dummy08} Carney, B. W., Latham, D. W., 1987, AJ, 93, 116
\bibitem[dummy]{dummy09} Churchill, C. W., Allen, S. L., 1995, PASP, 107, 193
\bibitem[dummy]{dummy11} Cram, L. E., 1983, PASA, 5, 152
\bibitem[dummy]{dummy12} Cuntz, M., Rammacher, W., Ulmschneider, P., 1994, ApJ, 432, 690
\bibitem[dummy]{dummy13} Duncan, D. K., et al., 1991, ApJS, 76, 383
\bibitem[dummy]{dummy14} Durney, B. R., De Young, D. S., Roxburgh, I. W., 1993, Sol. Phys., 145, 207
\bibitem[dummy]{dummy15} Giclas, H. L., Burnham, R., Jr., Thomas,
N. G., 1971, Lowell Proper Motion Survey, Northern Hemisphere (Lowell
Observatory, Flagstaff)
\bibitem[dummy]{dummy16} Hartmann, L. W., Noyes, R. W., 1987, ARAA, 25, 271
\bibitem[dummy]{dummy17} Laird, J. B., Carney, B. W., Latham, D. W., 1988, AJ, 95, 1843
\bibitem[dummy]{dummy18} Linsky, J. L., 1980, ARAA, 18, 439
\bibitem[dummy]{dummy19} Linsky, J. L., Worden, S. P., McClintock, W., Robertson, R. M., 1979,
     ApJS, 41, 47
\bibitem[dummy]{dummy21} Middelkoop, F. 1981, A\&A, 101, 295
\bibitem[dummy]{dummy22} Noyes, R. W., Hartmann, L. W., Baliunas, S. L., Duncan, D. K., Vaughan,
     A. H., 1984, ApJ, 279, 763
\bibitem[dummy]{dummy23} Oranje, B. J. 1983, A\&A, 124, 43
\bibitem[dummy]{dummy24} Oranje, B. J., Zwaan, C., 1985, A\&A, 147, 265
\bibitem[dummy]{dummy25} Peterson, R. C., Schrijver, C. J., 1997, ApJ, 480, L47
\bibitem[dummy]{dummy26} Rutten, R. J., Uitenbroek, H. 1991, Sol. Phys., 134, 15
\bibitem[dummy]{dummy27} Schrijver, C. J., 1987a, A\&A, 172, 111
\bibitem[dummy]{dummy28} Schrijver, C. J., 1987b, in {\it Cool Stars, Stellar Systems, and the Sun},
     eds. J. L. Linsky \& R. E. Stencel (Berlin: Springer-Verlag), p. 135
\bibitem[dummy]{dummy29} Schrijver, C., 1995, A\&AR, 6, 181 
\bibitem[dummy]{dummy30} Simon, T., Herbig, G. H., Boesgaard, A. M., 1985, ApJ, 293, 551
\bibitem[dummy]{dummy31} Skumanich, A., 1972, ApJ, 171, 565
\bibitem[dummy]{dummy32} White, O. R., Livingston, W. C., 1981, ApJ, 249, 798
\bibitem[dummy]{dummy33} Wilson, O. C., 1976, ApJ, 205, 823
\bibitem[dummy]{dummy34} Vaughan, A. H., Preston, G. W., 1980, PASP, 92, 385
\end{thebibliography}
\end{document}